# Dynamically encircling an exceptional point through phase-tracked closed-loop control

Sen Zhang[1], Yangyu Huang[1], Lei Yu[2], Kaixuan He[2], Ning Zhou[2], Dingbang Xiao[1], Xuezhong Wu[1], Franco Nori[3,4], Hui Jing[5*], Xin Zhou[1*]

[1]College of Intelligence Science and Technology, NUDT, Changsha 410073, China.

[2]East China Institute of Photo-Electronic IC, Bengbu 233042, China.

[3]Center for Quantum Computing (RQC), RIKEN, Wakoshi, Saitama, 351-0198, Japan.

[4]Department of Physics, University of Michigan, Ann Arbor, MI 48109-1040, USA.

[5]Key Laboratory of Low-Dimensional Quantum Structures and Quantum Control of Ministry of Education, Hunan Normal University, Changsha, 410081, China.

*Corresponding author. E-mail: jinghui73@gmail.com; zhouxin11@nudt.edu.cn.

**Abstract:** The intricate complex eigenvalues of non-Hermitian Hamiltonians manifest as Riemann surfaces in control parameter spaces. At the exceptional points (EPs), the degeneracy of both eigenvalues and eigenvectors introduces noteworthy topological features, particularly during the encirclement of the EPs. Traditional methods for probing the state information on the Riemann surfaces involve static measurements; however, realizing continuous encircling remains a formidable challenge due to non-adiabatic transitions that disrupt the transport paths. Here we propose an approach leveraging the phase-locked loop (PLL) technique to facilitate smooth, dynamic encircling of EPs while maintaining resonance. Our methodology strategically ties the excitation frequencies of steady states to their response phases, enabling controlled traversal along the Riemann surfaces of real eigenvalues. This study advances the concept of phase-tracked dynamical encircling and explores its practical implementation within a fully electrically controlled non-Hermitian microelectromechanical system, highlighting robust in-situ tunability and providing methods for exploring non-Hermitian topologies.

## Introduction

The complex eigenvalues of non-Hermitian Hamiltonians appear in the form of the roots of some complex functions, with their mapping over control parameter spaces giving rise to Riemann surfaces. At the pivotal branch points of these surfaces, termed exceptional points (EPs), both eigenvalues and eigenvectors undergo degeneracy. The unusual topology of the eigenvalue manifold and the skewed basis in the vicinity of the EP can exhibit remarkable nontrivial behaviors that hold great potential for various applications [1-5]. Of particular significance are the topological properties observed when EPs are encircled [6-9. Commonly employed methods to probe eigenvalue and eigenstate information on Riemann surfaces involve discrete measurements of the system's static properties at various parametric conditions along a closed loop [10-15]. Techniques such as stroboscopic measurement of limited eigenmode profiles on the loop allow for the determination of the Berry phase in non-Hermitian systems [10-13], while braiding of eigenvalues has been demonstrated through the analysis of steady-state spectra [14,15]. However, the primary utility of discrete measurements resides in the characterization of topological properties within non-Hermitian systems, which inherently constrains the practical applications of topological properties. Moreover, such measurements require spectral analysis at each discrete sampling point, introducing considerable computational redundancy. Dynamic measurements provide a methodology to mitigate the aforementioned challenges, while achieving genuine continuous braiding or real-time accumulation of the Berry phase through dynamic execution of smooth encircling remains a formidable challenge, especially since the dynamical encircling of EPs often encounters non-adiabatic transitions [16-17]. In this case, the transport path abruptly leaves the Riemann surfaces from a higher-loss (lower-gain) mode to a lower-loss (higher-gain) one, as elucidated by the Stokes phenomenon of asymptotics [18-19]. It remains an open question to dynamically and continuously transport that maintains presence on the Riemann surfaces for all encircling paths.

To address this challenge, it is essential to sustain resonance even when the system changes with time. The phase-locked loop (PLL) technique emerges as a promising solution 20, having successfully maintained resonant oscillations in billions of electronic devices, including frequency reference oscillators in smartphones 21 and inertial sensors in automobiles 22. By locking the phase of a resonator relative to external actuation, PLL can preserve a stable oscillation condition, even in the face of environmental perturbations that might otherwise disrupt its resonant frequency.

While non-Hermitian systems have been explored across various domains, most realizations have predominantly utilized optical or photonic controls [1-5]. The exploration of non-Hermicity within electrically tunable integrated mechanical systems—renowned for their superior in-situ controllability [23-32] and wide-ranging applications [21,22,34,35]—has proven elusive.

In this study, we leverage the PLL technique to propose a method for smoothly encircling EPs in a dynamic and continuous manner, ensuring that the transport path adheres closely to the eigenfrequency Riemann surface. This innovative approach emphasizes the closed-loop control of steady-state responses rather than their transient behavior. By controlling excitation frequencies through tracking response phases, we can emulate and trace out arbitrary trajectories on the eigenfrequency Riemann surface, opening avenues for exploring and exploiting the eigenvalue and eigenstate topological phenomena. Furthermore, we introduce a fully electrically controlled non-Hermitian microelectromechanical system (MEMS) integrated onto a single chip. Leveraging the excellent in-situ controllability and tunability of these systems, we experimentally showcase the efficacy of our phase-tracked dynamical encircling method within the non-Hermitian MEMS resonator.

## Results

### Concept

Here, we consider a second-order non-Hermitian system (highlighted by colored parts in Fig. 1a) described by the following Hamiltonian,

$$\mathbf{H} = \begin{pmatrix} \Omega_1 - \mathrm{i}\frac{\gamma_1}{2} & g \\ g & \Omega_2 - \mathrm{i}\frac{\gamma_2}{2} \end{pmatrix}, \tag{1}$$

where $\Omega_{1,2}$ and $\gamma_{1,2}$ represent the natural frequencies and damping rates of the two coupled modes. The coupling rate $g$ and the mode frequency detuning $\Delta = \Omega_2 - \Omega_1$ serve as the two controlling parameters of this system. Moreover, non-Hermiticity arises from the condition $\gamma = \gamma_2 - \gamma_1 \neq 0$. To actuate the system, an external force of the form $F\cos(\omega_\mathrm{d} t)$ is applied to mode 1. The responding displacement of this actuated mode is described by $A\cos(\omega_\mathrm{d}\, t + \theta)$, where the amplitude $A$ and phase $\theta$ are functions of excitation frequency $\omega_\mathrm{d}$. A typical response of the coupled system is shown in Fig. 1b. The phase response $\theta$ can be mapped to the excitation frequency $\omega_\mathrm{d}$ through $\theta = -\mathrm{Arg}[\chi_1(\omega_\mathrm{d})]$, where $\chi_1(\omega_\mathrm{d})$ is the susceptibility of the driven mode. In this study, the system is operated in the weak-coupling region, specifically $2g < \max(\gamma_1, \gamma_2)$, to ensure that the mapping remains injective, otherwise, bifurcation would take place 31. This mapping can be harnessed to establish a closed-loop oscillation through PLL techniques [31,36], enabling the control of the oscillation frequency to maintain the response phase $\theta$ at desired values.

We focus on the eigenfrequencies $\mathrm{Re}(\lambda_\pm)$ of the non-Hermitian system, where $\lambda_\pm$ represent the complex eigenvalues of the Hamiltonian. The eigenmode spectra are illustrated by the purple and green dashed Lorentzian curves in Fig. 1b. The eigenfrequencies $\mathrm{Re}(\lambda_\pm)$, indicated by dot-dashed lines aligned with the peaks of the Lorentzian curves, are mapped to two values of the response phase $\theta[\mathrm{Re}(\lambda_\pm)] = -\mathrm{Arg}\{\chi_1[\mathrm{Re}(\lambda_\pm)]\}$. Furthermore, eigenfrequencies $\mathrm{Re}(\lambda_\pm)$ as functions of the coupling $g$ and detuning $\Delta$ is depicted in Fig. 1c, and their mapping to $\theta[\mathrm{Re}(\lambda_\pm)]$ is presented in Fig. 1d.

We dynamically traverse a closed path in the parameter space surrounding the EP by continuously varying the coupling $g(t)$ and mode detuning $\Delta(t)$ over time, while simultaneously tracking the corresponding phase $\theta\{\mathrm{Re}[\lambda_\pm(t)]\}$ (represented by the red and blue curves in Fig. 1d) using a PLL, as illustrated in Fig. 1a. This process enables the attainment of a self-sustained closed-loop oscillation. The evolution of the oscillation frequency closely emulates a fully adiabatic process on the eigenfrequency Riemann surface, as evidenced by the red and blue curves in Fig. 1c. These controlled loops facilitate the permutation of excitations between the two sheets, irrespective of the direction of encircling.

**<u>On-chip non-Hermitian system</u>**

To facilitate the observation of the smooth encircling, we construct an on-chip non-Hermitian system utilizing a silicon-based MEMS disk resonator, as illustrated in Fig. 2a. The resonator devices are fabricated through bonding, deep reactive ion etching (DRIE), and chemical mechanical polishing (CMP) processes on two silicon-on-insulator (SOI) wafer. After processing, the wafer is diced using stealth laser cutting technology to obtain individual devices.

As shown in Fig. 2b, this system features a pair of six-node standing-wave modes, with natural frequencies $\omega_1/2\pi = 50{,}468.68$ Hz and $\omega_2/2\pi = 51{,}007.86$ Hz. The characteristic structural thicknesses of the deforming beams corresponding to the two modes are deliberately designed to differ, leading to variations in their thermal-elastic dissipations and overall quality factors (See Methods). This tailored design enables the engineering of the damping rates for the two modes, resulting in $\gamma_1 = 2\pi \times 676$ mHz and $\gamma_2 = 2\pi \times 904$ mHz, yielding a considerable difference of $\gamma = 2\pi \times 228$ mHz.

As shown in Fig. 2c, the two six-node modes of this resonator are dynamically coupled through the application of an anti-Stokes parametric pump $V_\mathrm{p}\cos(\omega_\mathrm{p} t)$, which induces a Floquet modulation of the intermodal coupling via electrostatic effects 29 (See Methods). Here, $V_\mathrm{p}$ represents the pump strength, and $\omega_\mathrm{p}$ is the pump frequency, closely tuned to the mode frequency mismatch $\Delta\omega = \omega_2 - \omega_1$. The dynamic coupling can be regulated in real time by tuning the pump $V_\mathrm{p}\cos(\omega_\mathrm{p} t)$, showcasing remarkable in-situ controllability.

Mode 1 is coherently coupled to the first harmonic idler wave of mode 2. In the rotating frame of pump frequency, the system is described by an effective non-Hermitian Hamiltonian of the form in equation (1), where $\Omega_1 = \omega_1 - \kappa V_\mathrm{p}^2/(8\omega_1)$, $\Omega_2 = \omega_2 - \kappa V_\mathrm{p}^2/(8\omega_2) - \omega_\mathrm{p} = \omega_1 - \kappa V_\mathrm{p}^2/(8\omega_2) - \delta_\mathrm{p}$, and $g = \kappa V_0 V_\mathrm{p}/(4\omega_1)$. Here, $\kappa$ is the electrostatic tuning coefficient, measured to be $70{,}186$ N m$^{-1}$ kg$^{-1}$ V$^{-2}$, while $V_0 = 40$ V is a static bias voltage applied to the resonator body. The term $\delta_\mathrm{p} = \omega_\mathrm{p} - \Delta\omega$ denotes the pump detuning. The experimentally feasible control parameters for this non-Hermitian system are $V_\mathrm{p}$ and $\delta_\mathrm{p}$, which govern the coupling strength and effective degeneracy, respectively. It is worth noticing that $V_\mathrm{p}$ may also influence degeneracy by an amount of $\kappa V_\mathrm{p}^2/(8\omega_1) - \kappa V_\mathrm{p}^2/(8\omega_2)$, though this effect is negligible within the measurement range of this study.

To construct the Riemann surfaces of the eigenvalues, we actuate mode 1 and measure its open-loop steady-state frequency responses as functions of the excitation frequency $\omega_\mathrm{d}$ across varying values of $V_\mathrm{p}$ and $\delta_\mathrm{p}$. The resonant frequencies $\mathrm{Re}(\lambda_\pm)$ and linewidths $-2\mathrm{Im}(\lambda_\pm)$ of the normal modes are determined by fitting the steady-state frequency-response spectra to the scaled susceptibility of mode 1 (Details see Methods). The results are shown in Fig. 2d-e, where the points represent experimental data, and the surfaces correspond to theoretical results calculated using the best-fit parameters. Additionally, the red points, obtained near degeneracy ($\delta_{\mathrm{p}} \approx 0$), illustrate the process of the parity-time phase transition. The transition point, located at $(V_{\mathrm{p}}, \delta_{\mathrm{p}}/2\pi) \approx (162\ \mathrm{mV}, 0\ \mathrm{Hz})$, corresponds to the EP, serving as the branch point of the Riemann surfaces.

**Smooth encircling of EP**

To achieve smooth encircling of the EP on the eigenfrequency $\mathrm{Re}(\lambda_{\pm})$ Riemann surface, we implement an adaptive PLL designed to track the response phase $\theta$ of the driven mode to the time-varying values of $\theta\{\mathrm{Re}[\lambda_{\pm}(t)]\}$ along the encircling paths in Fig. 1d. As shown in Fig. 3a, the essence of this control loop lies in the introduction of a controlled phase shift before the reference signal is put into the phase detection module. The light blue phase-shifting block is capable of updating the phase shift $\phi(t)$ in real-time. The phase detector measures the relative phase $(\theta - \phi)$ between the response signal of the driven mode $A\cos(\omega_{\mathrm{d}} t + \theta)$ and the phase-shifted reference signal $\cos(\omega_{\mathrm{d}} t + \phi)$. A proportional-integral-derivative (PID) controller is then employed to adjust the reference frequency $\omega_{\mathrm{d}}$, ensuring that the relative phase $\theta - \phi$ remains locked at a constant setpoint value $\theta_0$. The controlled reference signal is subsequently scaled to generate the appropriate driving force. Through this adaptive PLL control, the response phase is effectively tracked to follow the time-varying set values $\theta = \theta_0 + \phi(t)$.

The parametric loop for encircling is controlled by the strength $V_{\mathrm{p}}$ and the detuning $\delta_{\mathrm{p}}$, of the pump that is produced by a programmable signal generator. The base point for the topological encircling is selected as $(V_{\mathrm{p}}, \delta_{\mathrm{p}}/2\pi) = (0.5\ \mathrm{V}, 0.3\ \mathrm{Hz})$, which is located in the PT-symmetric phase. In this study, we implement a clockwise (CW) rectangular parametric loop defined by the sequence $[(V_p, \delta_p/2\pi)] = [(0.5\ \mathrm{V}, 0.3\ \mathrm{Hz}), (0.1\ \mathrm{V}, 0.3\ \mathrm{Hz}), (0.1\ \mathrm{V}, -0.3\ \mathrm{Hz}), (0.5\ \mathrm{V}, -0.3\ \mathrm{Hz}), (0.5\ \mathrm{V}, 0.3\ \mathrm{Hz})]$, as well as its reverse loop. Both loops enclose the EP.

The trajectory of the smooth encircling is defined by the parameters $(V_p(t), \delta_p(t), \theta\{\mathrm{Re}[\lambda_\pm(t)]\})$, where $V_p(t)$ and $\delta_p(t)$ represent the time evolution along the parametric loop, and $\theta\{\mathrm{Re}[\lambda_\pm(t)]\}$ is the corresponding dynamic evolution of the locking phase. The phase $\theta\{\mathrm{Re}[\lambda_\pm(t)]\}$ for each loop is predetermined by calculating $-\mathrm{Arg}\{\chi_1[\mathrm{Re}(\lambda_\pm)]\}$ using the best-fit parameters of the non-Hermitian system derived from Fig. 2d-e. In practice, the test circuitry can introduce a constant phase shift of $-65°$ to the responding phase $\theta$. This phase shift is compensated before the theoretical $\theta\{\mathrm{Re}[\lambda_\pm(t)]\}$ is put into the adaptive PLL (see Methods).

This process requires simultaneous control of the pump signal generator and the adaptive PLL (Details see Methods). The starting point location is determined on the high- or low-frequency sheets of the Riemann surface by setting $\theta_0 = \theta\{\mathrm{Re}[\lambda_+(0)]\}$ or $\theta\{\mathrm{Re}[\lambda_-(0)]\}$, respectively. Throughout the smooth encircling process, the time-varying phase shift is defined as $\phi(t) = \theta\{\mathrm{Re}[\lambda_\pm(t)]\} - \theta\{\mathrm{Re}[\lambda_\pm(0)]\}$. The sign of $\pm$ is switched upon crossing the branch cut of the Riemann surface. In practice, the continuous evolution of $\phi(t)$ is implemented by successively inputting 2,001 predefined pinpoint values into the phase shifter in a time series synchronized with the pump generation.

We conduct four encircling processes-clockwise (CW) and counterclockwise (CCW)-starting from the high and low-frequency sheets of the Riemann surface, respectively, all of which enclose the EP. Fig. 3b-d present the implementation of a CW encircling starting from the high-frequency sheet of the Riemann surface. As shown in Fig. 3b, the response phase $\theta$ (red solid curve) is closely tracked to $\theta\{\mathrm{Re}[\lambda_\pm(t)]\}$ (gray solid curve). This is ensured by two key factors: the relative phase $\theta - \phi$ (green solid curve) is locked to $\theta_0 = \theta\{\mathrm{Re}[\lambda_+(0)]\}$ (black dot-dashed curve), and the phase shift $\phi(t)$ evolves in accordance with $\theta\{\mathrm{Re}[\lambda_{\pm}(t)]\} - \theta\{\mathrm{Re}[\lambda_{+}(0)]\}$ (blue gradients). In the meantime, the control parameters $V_{\mathrm{p}}(t)$ and $\delta_{\mathrm{p}}(t)$ trace out a CW loop, as shown in Fig. 3c. Additionally, the continuous output of the controlled closed-loop oscillation frequency is processed to yield its moving average (red thick curve), which aligns well with the theoretical value on the $\mathrm{Re}[\lambda_{\pm}(t)]$ Riemann surface (gray solid curve), as shown in Fig. 3d. The error band represents the rolling standard deviation of the measured frequency.

The results of the four encircling processes are presented in Fig. 4. The measured oscillation frequency (moving averaged value) and the corresponding tracked phase of the smooth encircling processes illustrated on the $\mathrm{Re}[\lambda_{\pm}(t)]$ and $\theta\{\mathrm{Re}[\lambda_{\pm}(t)]\}$ surfaces are shown in Fig. 4a and b, respectively.

The rectangular path above varies only one parameter at any time. We also constructed a circular path where two parameters are varied simultaneously to better demonstrate the feasibility of smoothly encircling EP through phase-tracked closed-loop control (see Methods). The full results of encircling EP with a circular path are shown in Supplementary Figure 1.

The full results of Fig. 4 and Supplementary Figure 1 depict trajectories for the CW encircling starting from the high-frequency sheet (red trajectory), the CW encircling starting from the low-frequency sheet (blue trajectory), the CCW encircling starting from the high-frequency sheet (purple trajectory), and the CCW encircling starting from the low-frequency sheet (green trajectory). These findings robustly demonstrate that a phase-tracked dynamic encircling of the EP can permute excitations between the two sheets, regardless of the direction of encircling.

To demonstrate that the encircling of EP is not only independent of the encircling direction, but also independent of the position of the starting point, we also conducted an encircling path starting at the PT-broken phase. The base point for the encircling is selected as $\left(V_{\mathrm{p}}, \delta_{\mathrm{p}}/2\pi\right) =$ $(0.1\ \mathrm{V}, 0.3\ \mathrm{Hz})$, which is located in the PT-broken phase. In the CW encircling path, the control parameters vary linearly by the sequence: $\left[\left(V_{\mathrm{p}}, \delta_{\mathrm{p}}/2\pi\right)\right] = [(0.1\ \mathrm{V}, 0.3\ \mathrm{Hz}), (0.1\ \mathrm{V}, -0.3\ \mathrm{Hz}), (0.5\ \mathrm{V}, -0.3\ \mathrm{Hz}), (0.5\ \mathrm{V}, 0.3\ \mathrm{Hz}), (0.1\ \mathrm{V}, 0.3\ \mathrm{Hz})]$, while the parameters follow the reverse sequence in the CCW encircling path. The results of the encircling processes are shown in Fig. 5. The dynamical encircling of EP initiated from the PT-broken phase does not manifest chiral behavior, which demonstrates that the phase-tracked dynamic encircling of the EP enables adiabatic and continuous dynamical evolution.

## Conclusions

In summary, our study presents a robust method for smoothly traversing the eigenfrequency Riemann surface of non-Hermitian systems. The state corresponding to the phase-tracked encircling trajectory is represented by the hybrid state described by $|\psi\rangle = \frac{-1}{(\lambda_{+}-\omega_{\mathrm{d}})(\lambda_{-}-\omega_{\mathrm{d}})}\left(\omega_{\mathrm{d}} - \Omega_2 + \mathrm{i}\frac{\gamma_2}{2}, g\right)^{\mathrm{T}}$ in the basis spanned by modes 1 and 2, with $\omega_{\mathrm{d}} = \mathrm{Re}[\lambda_{\pm}]$. By further measuring the instantaneous phase-tracked hybrid state information, one can extract the information of the instantaneous Hamiltonian $\mathbf{H}$, recover the imaginary part of the instantaneous eigenvalue $\mathrm{Im}[\lambda_{\pm}]$, and reconstruct the instantaneous eigenstates $|v_{\pm}\rangle = e^{\mathrm{i}\alpha}\left(\lambda_{\pm} - \Omega_2 + \mathrm{i}\frac{\gamma_2}{2}, g\right)^{\mathrm{T}}$ where $\alpha$ is an arbitrary phase factor. In this dynamically coupled device, we currently face limitations in measuring the phase information of mode 2 due to its strong dependence on the uncontrolled pump phase. We aim to address this challenge in future work by constructing degenerate non-Hermitian MEMS oscillators coupled through ordinary coherent coupling. This strategy holds great promise for realizing genuine continuous eigenvalue braiding and observing real-time Berry phase accumulation.

More broadly, this work establishes a methodology for achieving robust topological control in non-Hermitian oscillator systems. Beyond emulating adiabatic evolution, our proposed method can also induce non-adiabatic transitions by intentionally modulating the resonant phase switch between high-frequency and low-frequency sheets (see Methods and Supplementary Figure 2), providing a platform for exploring the non-Hermitian physical properties of non-adiabatic transitions.

# Methods

## Device design

One effective approach for constructing a non-Hermitian MEMS resonator involves engineering a pair of near-degenerate modes with distinct damping rates, as demonstrated in the uniquely designed MEMS disk resonator of this study. The primary dissipation mechanism in the disk resonator is thermal-elastic damping (TED). By strategically regulating TED through structural design, it is possible to effectively manipulate the overall quality factor ($Q$) [1-39].

The TED arises from the coupling of the strain field to the temperature field 40. During instantaneous deformation, tensile strain induces cooling, while compressive strain leads to heating, resulting in local temperature gradients. The relaxation of this thermal nonequilibrium back to equilibrium causes an irreversible flow of heat, breaking the conservation of kinetic and potential energy in the vibrating resonator. Zener's standard model for TED is expressed as:

$$Q_{\mathrm{TED}} = \frac{C}{E\alpha^2 T_0}\left(\frac{1}{\omega_{\mathrm{m}}\tau_{\mathrm{Z}}} + \omega_{\mathrm{m}}\tau_{\mathrm{Z}}\right), \qquad (2)$$

where $Q_{\mathrm{TED}}$ is the thermal-elastic quality factor, $C$ is the heat capacity of the silicon material, $E$ is Young's modulus, $\alpha$ is the coefficient of thermal expansion, $T_0$ is the equilibrium temperature, $\omega_{\mathrm{m}}$ is the angular frequency of mechanical vibration, and $\tau_{\mathrm{Z}}$ is the thermal relaxation time, defined as the time required for the system to return to equilibrium following the establishment of a local temperature gradient:

$$\tau_{\mathrm{Z}} = \frac{b^2}{\pi^2 \chi}. \tag{3}$$

Here, $\chi$ is the thermal diffusivity of the solid, and $b$ represents the width of the beam or the length of the heat flow path. In the flexible MEMS device studied, the vibration frequency $\omega_{\mathrm{m}}$ is significantly smaller than the rate of thermal relaxation $1/\tau_{\mathrm{Z}}$. Under these isothermal conditions, the term $\omega_{\mathrm{m}}\tau_{\mathrm{Z}}$ can be neglected in comparison to the $1/(\omega_{\mathrm{m}}\tau_{\mathrm{Z}})$ term in equation (2). Consequently, using a thinner beam or decreasing $\tau_{\mathrm{Z}}$ can yield a higher $Q_{\mathrm{TED}}$ [41-42].

To regulate TED, we propose a disk resonator with a diameter of 4 mm, featuring 12 circularly connected fan-shaped sectors composed of either concentrically spaced double-thin beams (width $b_1 = 9$ μm) or single-thick beams (width $b_2 = 13$ μm), as illustrated in Supplementary Figure 3a. The disk resonator is anchored at its center and is surrounded by 24 capacitive electrodes that facilitate the excitation, transduction, and tuning of the resonator. The width of the capacitive gap between the electrodes and resonator is 9 μm, and both the structure and electrodes have a height of 100 μm.

The deformations of the flexible beams give rise to a pair of six-node in-plane standing-wave modes, as depicted in Supplementary Figure 3b. The instantaneous temperature-deviation field resulting from the thermal-elastic coupling of the operational modes is shown in Supplementary Figure 3c. The TED processes predominantly occur in the nodal (N) sectors, as indicated by the deeper colors of the temperature-deviation fields in these sectors compared to the anti-nodal (A) sectors in Supplementary Figure 3c. In our innovative design, mode 1 incorporates thinner nodal beams ($b_1 < b_2$), which contribute to reduced TED and a higher thermal-elastic quality factor ($Q_{\mathrm{TED}}$). The simulated thermal-elastic quality factors for the operational modes are $Q_{\mathrm{TED},1} = 88{,}024$ and $Q_{\mathrm{TED},2} = 65{,}389$, as determined through TED simulations conducted using COMSOL.

The device is fabricated from <100> single-crystal silicon using a MEMS process as detailed in Supplementary Note 1. By experimentally measuring the decay signals of the two operational modes in the device, the quality factors were determined to be $Q_1 = 74{,}658$ and $Q_2 = 56{,}424$. The corresponding damping rates are then calculated as $\gamma_{1,2} = \omega_{1,2}/Q_{1,2}$, leading to a significant damping difference of $\Delta\gamma = 2\pi \times 228$ mHz.

While we have meticulously engineered the beam width differences to achieve the closest possible natural frequencies for the two modes, an uncompensatable frequency mismatch of $\Delta\omega = 2\pi \times 539.18$ Hz still arises in the measured device, attributable to the fabrication tolerances inherent in the MEMS manufacturing process. Future improvements in degeneracy may be attained by further refining the beam width differences.

**Experimental setup**

More details about the experimental setup are provided in Supplementary Figure 4a. The packaged MEMS device is integrated with signal processing circuitry on a printed circuit board. The disk resonator is differentially actuated to excite mode 1 by applying two anti-phase alternating signals generated by a lock-in amplifier (Zurich Instruments MFLI) to the respective anti-phase antinodal electrodes of mode 1. Throughout the experiment, the amplitude of the drive signal is maintained at a constant level. A constant bias voltage of $V_0 = 40$ V is applied to the resonator body, providing the electrostatic potential energy necessary for both actuation and tuning. Two anti-phase antinodal displacements of the two modes are differentially detected using two charge amplifiers. The resultant displacement signal is then amplified and fed into the lock-in amplifier. Additionally, an alternating pump signal generated by a programmable wave generator is applied to a set of nodal electrodes that are offset by $+15°$ compared to the actuation electrodes. The pump strength $V_{\mathrm{p}}(t)$ and frequency $\omega_{\mathrm{p}}(t)$ are programmed to vary gradually in order to trace out the encircling parametric loop.

The test circuitry can introduce an additional phase shift, denoted as $\varphi$ in the output displacement signal. By fitting, we find that $\varphi$ is of a constant value $-65°$. To address this, a phase shift block is incorporated to compensate for $\varphi$ prior to inputting the displacement signal into the adaptive PLL. The phase responses obtained from an open-loop frequency-sweeping measurement, both with and without phase compensation, are illustrated in Supplementary Figure 4b. After the implementation of phase compensation, the experimentally measured phase of the displacement signal coincides well with the theoretical values calculated based on $\theta = -\mathrm{Arg}(\chi_1)$.

During closed-loop operation, the mode-1 displacement signal $A\cos(\omega_{\mathrm{d}}\, t + \theta)$ is demodulated by the phase-shifted reference signal $\cos(\omega_{\mathrm{d}}\, t + \phi)$ within the phase detector. The relative phase $\theta - \phi$ is maintained at a predefined setpoint $\theta\{\mathrm{Re}[\lambda_{+}(0)]\}$. The phase shift $\phi(t)$ of the reference signal is continuously updated using 2,001 values of $\{\mathrm{Re}[\lambda_{\pm}(t)]\} - \theta\{\mathrm{Re}[\lambda_{\pm}(0)]\}$ along the encircling trajectory. This phase updating is facilitated through the application programming interface (API) of the lock-in amplifier. The wave generator and the lock-in amplifier are synchronized to enable simultaneous modulation of the parameters $V_{\mathrm{p}}(t)$, $\delta_{\mathrm{p}}(t)$, and $\phi(t)$.

**Dynamical coupling**

To coherently couple the two modes, a red-detuned parametric pump signal $V_{\mathrm{p}}\cos(\omega_{\mathrm{p}} t)$ with $\omega_{\mathrm{p}} \approx \Delta\omega$ is applied to the +15° off-axis electrodes, producing a modulation of the intermodal stiffness given by $\Delta_{\mathrm{p}} = \kappa\left\{V_0^2 - \left[V_0 - V_{\mathrm{p}}\cos\left(\omega_{\mathrm{p}}t\right)\right]^2\right\}$, where $\kappa$ is the electrostatic tuning coefficient. The higher-harmonic term $\kappa V_{\mathrm{p}}^2/2 \cdot \cos\left(2\omega_{\mathrm{p}}t\right)$ in $\Delta_{\mathrm{p}}$can be neglected, as it contributes only to higher-order coupling terms that are not considered in this study. Thus, the parametric pump can be expressed as:

$$\Delta_{\mathrm{p}} = 2\kappa V_0 V_{\mathrm{p}}\cos\left(\omega_{\mathrm{p}}t\right) - \frac{\kappa V_{\mathrm{p}}^2}{2}. \tag{4}$$

This formulation encompasses a Floquet modulation with frequency $\omega_{\mathrm{p}}$ alongside a small static tuning.

The dynamically modulated two-mode system can be represented as an order-reduced model, as illustrated in Supplementary Figure 5. The Newtonian equations of motion for this system are expressed as follows:

$$\begin{bmatrix}\ddot{x}\\ \ddot{y}\end{bmatrix} + \begin{bmatrix}\gamma_1 & 0\\ 0 & \gamma_2\end{bmatrix}\begin{bmatrix}\dot{x}\\ \dot{y}\end{bmatrix} + \begin{bmatrix}\omega_1^2 + \Delta_{\mathrm{p}}/2 & \Delta_{\mathrm{p}}/2\\ \Delta_{\mathrm{p}}/2 & \omega_2^2 + \Delta_{\mathrm{p}}/2\end{bmatrix}\begin{bmatrix}x\\ y\end{bmatrix} = \begin{bmatrix}F\cos(\omega_{\mathrm{d}}t)/m\\ 0\end{bmatrix}, \tag{5}$$

where $x$ and $y$ denote the displacements of modes 1 and 2, respectively, and $m$ is the effective mass of mode 1. In the rotating frame at frequency $\omega_{\mathrm{d}}$ for mode 1 and the first idler wave frame at frequency $\omega_{\mathrm{d}} + \omega_{\mathrm{p}}$ for mode 2, the equations of motion can be simplified (details see Supplementary Note 2) as:

$$\mathrm{i}\begin{bmatrix}\dot{A}_0\\ \dot{B}_1\end{bmatrix} = \begin{bmatrix}\Omega_1 - \omega_{\mathrm{d}} - \mathrm{i}\gamma_1/2 & g\\ g & \Omega_2 - \omega_{\mathrm{d}} - \mathrm{i}\gamma_2/2\end{bmatrix}\begin{bmatrix}A_0\\ B_1\end{bmatrix} - \begin{bmatrix}f\\ 0\end{bmatrix}. \tag{6}$$

Here, $A_0$ represents the complex amplitude of mode 1, $B_1$ is the complex amplitude of the first idler wave of mode 2, $\Omega_1 = \omega_1 - \kappa V_{\mathrm{p}}^2/(8\omega_1)$, $\Omega_2 = \omega_2 - \kappa V_{\mathrm{p}}^2/(8\omega_2) - \omega_{\mathrm{p}} = \omega_1 - \kappa V_{\mathrm{p}}^2/(8\omega_2) - \delta_{\mathrm{p}}$, $g = \kappa V_0 V_{\mathrm{p}}/(4\omega_1)$, and $f = F/(4m\omega_1)$.

We then transform equ. (6) from the fast driving rotating frames to the slow Floquet frame by writing $x_0 = \frac{1}{2}A_0 \exp(-\mathrm{i}\omega_\mathrm{d}t) + \mathrm{c.c.}$ and $y_1 = \frac{1}{2}B_1 \exp(-\mathrm{i}\omega_\mathrm{d}t) + \mathrm{c.c.}$, where "c.c." signifies complex conjugation. The equations of motion can be further simplified into the following first-order complex differential equations (details see Supplementary Note 2):

$$\mathrm{i}\begin{bmatrix}\dot{x}_0\\ \dot{y}_1\end{bmatrix} = \begin{bmatrix}\Omega_1 - \mathrm{i}\gamma_1/2 & g\\ g & \Omega_2 - \mathrm{i}\gamma_2/2\end{bmatrix}\begin{bmatrix}x_0\\ y_1\end{bmatrix} - \begin{bmatrix}f\cos(\omega_\mathrm{d}t)\\ 0\end{bmatrix}. \tag{7}$$

The dynamical matrix $\begin{bmatrix}\Omega_1 - \mathrm{i}\gamma_1/2 & g\\ g & \Omega_2 - \mathrm{i}\gamma_2/2\end{bmatrix}$ describes the free evolution of the system and represents the non-Hermitian effective Hamiltonian. It is also possible to derive the Schrödinger-type equations of motion that are identical to equ. (7) quantum mechanically. However, since the system is purely classical, we employ the classical approach detailed above.

**Spectral fitting**

The open-loop responding displacement $A\cos(\omega_\mathrm{d}\,t + \theta)$ of mode 1 is recorded by the sweeper block of the lock-in amplifier. The amplitude $A$ and phase $\theta$ relative to the driving force are extracted through homodyne measurements.

The theoretical frequency responses of mode 1 can be calculated by applying the steady-state condition $\dot{A}_0 = \dot{B}_1 = 0$ in equ. (6). This results in $A_0 = f\chi_1(\omega_\mathrm{d})$, where the mechanical susceptibility of mode 1 is expressed as:

$$\chi_1(\omega_\mathrm{d}) = \frac{\Omega_2 - \omega_\mathrm{d} - \mathrm{i}\gamma_2/2}{(\omega_\mathrm{d} - \lambda_+)(\omega_\mathrm{d} - \lambda_-)}, \tag{8}$$

where

$$\lambda_\pm = \frac{\Omega_1 + \Omega_2}{2} - \mathrm{i}\frac{\gamma_1 + \gamma_2}{4} \pm \sqrt{\left(\frac{\Omega_2 - \Omega_1}{2} - \mathrm{i}\frac{\gamma_2 - \gamma_1}{4}\right)^2 + g^2} \tag{9}$$

are the eigenvalues of the Hamiltonian.

Ideally, $A = f|\chi_1(\omega_\mathrm{d})|$, and $\theta = -\mathrm{Arg}[\chi_1(\omega_\mathrm{d})]$. However, in practice, the circuitry will cause additional phase shift $\varphi$, and the input signal can also cause a small complex feedthrough $b \in \mathbb{C}$ signal to the output port. Consequently, we can fit the experimental data $A\exp(-\mathrm{i}\theta)$ using the following fitting function:

$$\mathcal{F}(\omega_\mathrm{d}) = f\chi_1(\omega_\mathrm{d})\mathrm{e}^{\mathrm{i}\varphi} + b. \tag{10}$$

To extract the eigenvalues from the steady-state frequency responses, the in-phase component $A\cos\theta$ and the quadrature component $A\sin\theta$ of the experimental data are fitted to the real part $\mathrm{Re}[\mathcal{F}(\omega_\mathrm{d})]$ and the negative imaginary part $-\mathrm{Im}[\mathcal{F}(\omega_\mathrm{d})]$ of the fitting function, respectively. This procedure yields the complex eigenvalues $\lambda_\pm$ of the non-Hermitian Hamiltonian. Supplementary Figure 6 presents fitting results for varying values of $V_\mathrm{p}$ while maintaining a constant $\delta_\mathrm{p}/2\pi = 0$ Hz. The eigenvalues obtained from these fittings are indicated by the red points in Fig. 2c. By extending this fitting process across a broader range of parametric conditions, we are able to populate the full Riemann surfaces depicted in Fig. 2d-e.

**Smoothly encircling of EP with a circular trajectory**

In the main text of this paper we have conducted a rectangular trajectory to dynamically encircle the EP, where only one parameter is varied at any time. However, there are many complex encircling trajectories in the scene of EP topological research. In these trajectories, the two control parameters of the system will change simultaneously. To further illustrate the feasibility of the control scheme proposed in this paper in the scenario where the control parameters change at the same time, here we construct a circular encircling trajectory, where the pump strength $V_\mathrm{p}(t)$ varies cosinusoidally with time ($V_\mathrm{p}(t) = 0.3 + 0.2\cos(\frac{\pi}{30}t)(\mathrm{V})$), and the pump frequency detuning $\delta_\mathrm{p}(t)$ varies sinusoidally ($\delta_\mathrm{p}(t) = 2\pi \cdot 0.2\sin(\frac{\pi}{30}t)(\mathrm{Hz})$).

We conducted four encircling experiments starting from the high-frequency sheet and low-frequency sheet of the Riemann surface, respectively, along both clockwise (CW) and counterclockwise (CCW) directions. As can be seen in Supplementary Figure 1, the oscillation frequency and the corresponding tracked phase of the system evolve smoothly along the Riemann surfaces of the real eigenvalues and the mapped phase respectively, and the end point of the evolution is independent of the circling direction, which demonstrate that it is also applicable to the encircling trajectory with two control parameters changing simultaneously.

**Non-adiabatic transitions through phase-tracked closed-loop control**

Different from Hermitian system, non-Hermitian system exhibits unique physical properties, one of which is the non-adiabatic transitions. Due to the Stokes phenomenon of asymptotics, the adiabatic evolution process of dynamically encircling EPs will be disrupted, leading to non-adiabatic transitions. Specifically, the system transitions from a high-dissipation mode to a low-dissipation mode on the Riemann surface, resulting in the final state of the encircling depending solely on the direction and being independent of the initial state. In previous studies, non-adiabatic transitions often occur only when the system is in a high-dissipation state. While in this section, we will demonstrate that non-adiabatic transitions can be induced in a low-dissipation state through phase-tracked closed-loop control.

As mentioned in the main text, by employing a phase-locked loop (PLL) to synchronize the resonant phase with the eigenfrequency-mapped phase, the system can sustain stable oscillation at any point on the eigenfrequency Riemann surface. Now we consider a closed loop path in the parameter space that encloses an EP. Along this path, the system possesses two eigenstates, with the high-frequency ($\mathrm{Re}(\lambda_+)$) and low-frequency ($\mathrm{Re}(\lambda_-)$) eigenstates corresponding to their respective resonant phases ($\theta\{\mathrm{Re}[\lambda_\pm(t)]\}$). By deliberately controlling the PLL to switch its tracking between the resonant phases of the high-frequency and low-frequency eigenstates, conversion between the high-frequency sheet and low-frequency sheet at any position can be realized, which enables the generation of non-adiabatic transition.

We conduct a CW rectangular parametric path defined by the sequence $\left[\left(V_\mathrm{p}, \delta_\mathrm{p}/2\pi\right)\right] = [(0.1\ \mathrm{V}, 0.3\ \mathrm{Hz}), (0.1\ \mathrm{V}, -0.3\ \mathrm{Hz}), (0.5\ \mathrm{V}, -0.3\ \mathrm{Hz}), (0.5\ \mathrm{V}, 0.3\ \mathrm{Hz}), (0.1\ \mathrm{V}, 0.3\ \mathrm{Hz})]$, as well as its reverse loop. In contrast to the smooth encircling presented in the main text, we deliberately induce a phase transition between the high- and low-frequency sheets in the dynamic encircling trajectory through low-dissipation region. As shown in Supplementary Figure 2, the oscillation frequency can be converted between the high-frequency sheet and the low-frequency sheet in the low-dissipation region, which indicates the occurrence of non-adiabatic transitions.

**Data availability**

The data that support the findings of this study are available within the main text and the extended data. Any other relevant data is available from the corresponding author upon request.

## Acknowledgments

This work is partly supported by the National Natural Science Foundation of China (NSFC) grant U21A20505 (D.X., X.Z., and X.W.), grant 11935006 (H.J.), the Hunan Provincial Major Sci-tech Program grant 2023zj1010 (H.J. and X.Z.), the Science and Technology Innovation Program of Hunan Province grant 2020RC4047 (H.J.). F.N. is supported in part by: the Japan Science and Technology Agency (JST) via the CREST Quantum Frontiers program Grant No. JPMJCR24I2, the Quantum Leap Flagship Program (Q-LEAP), and the Moonshot R&D Grant Number JPMJMS2061. This work is primarily supported by the National Key R&D Program of China (NKPs) grant 2022YFB3204901 (X.Z.).

## Author contributions

X.Z. conceived the idea and designed the research. S.Z., Y.H. and X.Z. conducted the experiments. X.Z. designed the device. L.Y., K.H., and N.Z. fabricated the device. S.Z., X.Z., D.X., and X.W. developed the test circuitry. X.Z., J.H., and F.N. conducted the theory. X.Z. wrote the manuscript with inputs from all authors. X.Z. and H.J jointly supervised the project.

## Competing interests

The authors declare no competing interests.

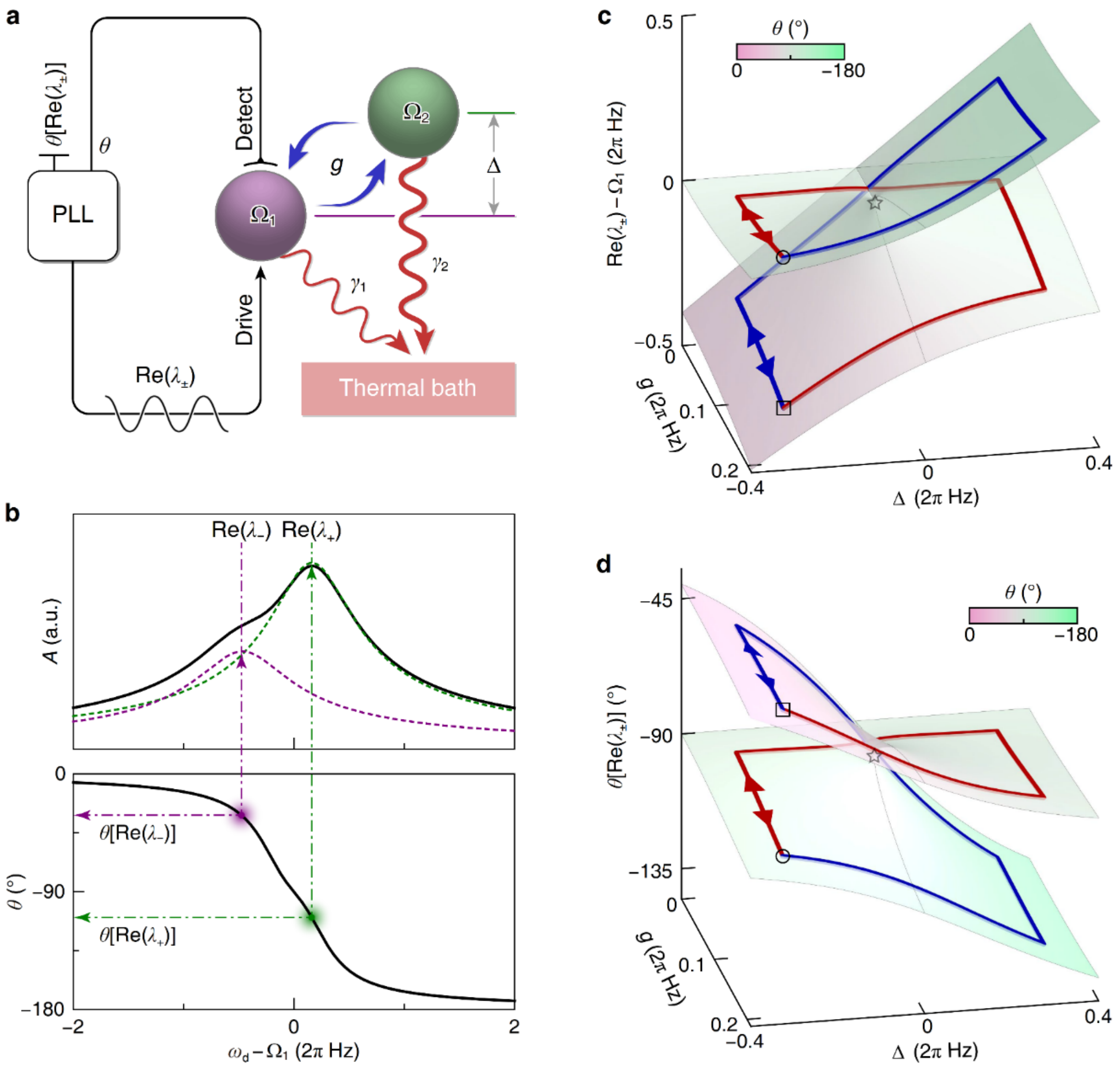

**Fig. 1. Concept of phase-tracked smooth encircling.** **a**, The non-Hermitian system comprises two coupled modes (labeled as $\Omega_1$ and $\Omega_2$ for their natural frequencies) with distinct damping rates. An adaptive phase-tracked closed-loop oscillation is established to trace the eigenfrequencies. PLL, phase-locked loop. **b**, The correspondence between the eigenfrequencies $\mathrm{Re}(\lambda_{\pm})$ and the response phase $\theta[\mathrm{Re}(\lambda_{\pm})]$ as illustrated in a typical steady-state spectrum. The purple and green dashed curves represent the eigenmode spectra. **c**,**d**, Eigenfrequencies $\mathrm{Re}(\lambda_{\pm})$ (**c**) and the corresponding response phase $\theta[\mathrm{Re}(\lambda_{\pm})]$ (**d**) are presented as functions of the coupling $g$ and detuning $\Delta$. The colors on the surfaces represent phase values. By smoothly following the red or blue directional paths in (**d**) using the phase-tracking technique depicted in (**a**), one can derive the corresponding red or blue dynamic evolutions of the eigenfrequencies in (**c**). The black circle and square mark the start/end points of the encircling trajectories on the high or low-frequency sheets, respectively, while the black star indicates the EP.

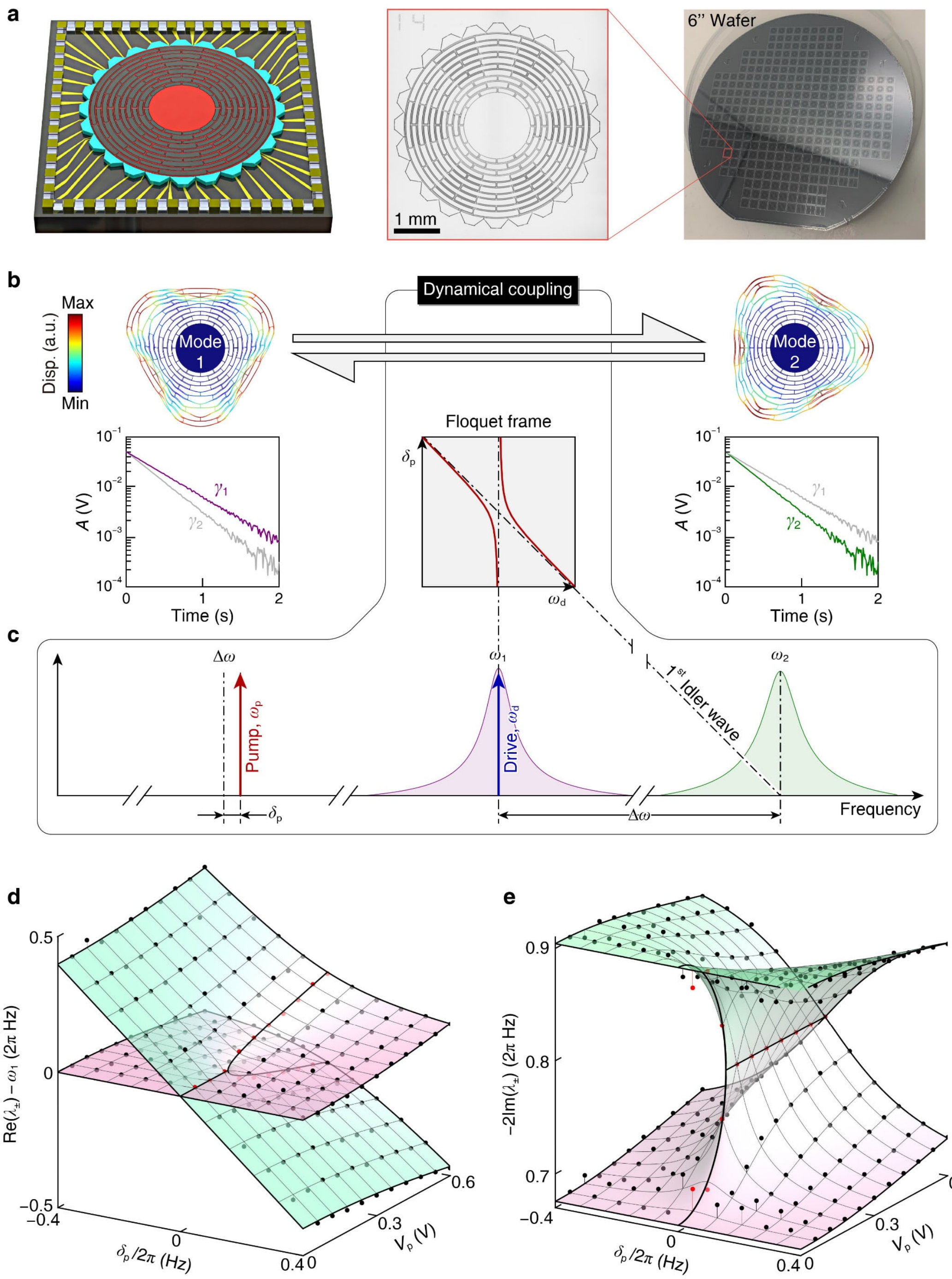

**Fig. 2. On-chip non-Hermitian microelectromechanical system.** **a**, The layout of this non-Hermitian resonator, where the left one shows the 3D model diagram, the right one shows a wafer containing 333 devices, and the middle one displays the microscope image of a single device (scale bar: 1mm). **b**, A pair of non-degenerate six-node standing-wave modes (labeled as 1 and 2) of a MEMS disk resonator are engineered to exhibit a considerable difference in damping rates. The insets present the simulated mode shapes and the experimental decay signals, respectively. At **c**, an anti-Stokes parametric pump with frequency $\omega_p \sim \Delta\omega$ is applied to establish a coherent dynamical coupling between mode 1 and the first harmonic idler wave of mode 2. The coupling strength is determined by the pump amplitude $V_p$. In the Floquet frame, the system is non-Hermitian. **d,e**, The resonant frequencies $\mathrm{Re}(\lambda_\pm)$ (**d**) and the linewidths $-2\mathrm{Im}(\lambda_\pm)$ (**e**) of the system are presented as functions of the pump strength $V_{\mathrm{p}}$ and pump detuning $\delta_{\mathrm{p}} = \omega_{\mathrm{p}} - \Delta\omega$. The points are obtained by fitting the experimental steady-state frequency-response spectra under varying conditions, while the surfaces represent theoretical values calculated using parameters derived from the fitting process. The red points illustrate the process of the parity-time phase transition while $\delta_{\mathrm{p}} \approx 0$. The colors on the surfaces correspond to the linewidth values.

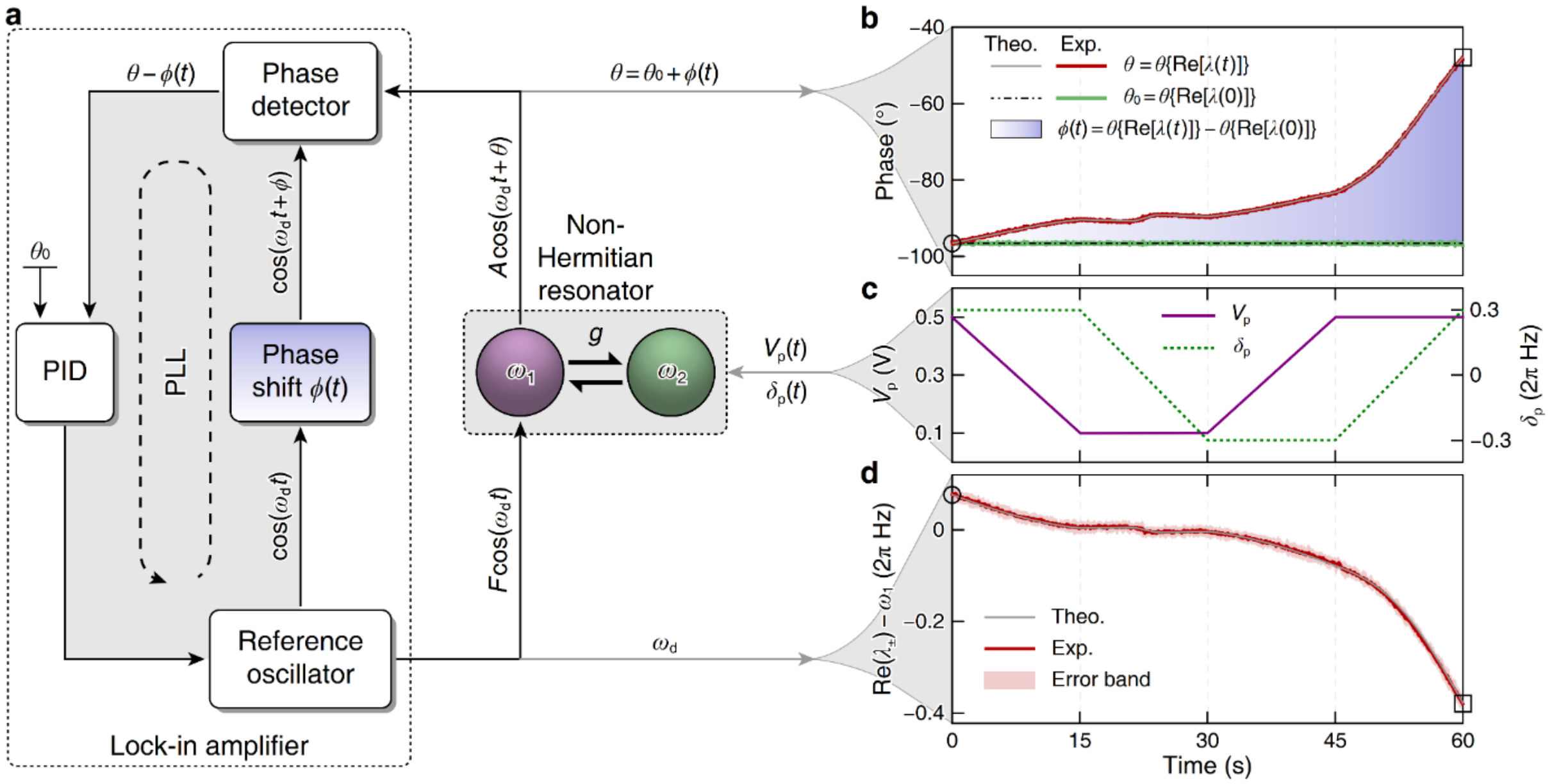

**Fig. 3. Setup for adaptive phase-tracking. a**, Mode 1 of the non-Hermitian resonator is actuated by a sinusoidal wave generated by a reference oscillator. The displacement of mode 1, which includes an additional response phase $\theta$, is demodulated using a parallel reference signal that is shifted by a time-varying phase $\phi(t)$ within a phase detector to yield the relative phase $\theta - \phi$. A PID controller adjusts the oscillator frequency $\omega_\mathrm{d}$ to maintain the relative phase $\theta - \phi$ at a constant setpoint value $\theta_0$. Overall, the response phase of mode 1 is tracked to a time-varying value, $\theta = \theta_0 + \phi(t)$. During the experiments, the pump strength $V_\mathrm{p}(t)$ and detuning $\delta_\mathrm{p}(t)$ are varied over time to trace out an encircling path. Simultaneously, we track $\theta$ to match the predefined value $\theta\{\mathrm{Re}[\lambda(t)]\}$ on the phase surface shown in Fig 1**d** by defining $\theta_0 = \theta\{\mathrm{Re}[\lambda_\pm(0)]\}$ and $\phi(t) = \theta\{\mathrm{Re}[\lambda_\pm(t)]\} - \theta\{\mathrm{Re}[\lambda_\pm(0)]\}$. **b-d**, Demonstration of a CW encircling process starting from the high-frequency sheet. Phase information (**b**), control parameters (**c**), and the corresponding oscillation frequency (**d**) of the encircling process. Theo., theoretical data. Exp., experimental results. The error band is the standard deviation. The colors of $\phi(t)$ in (**b**) correspond to the phase shift while encircling an EP.

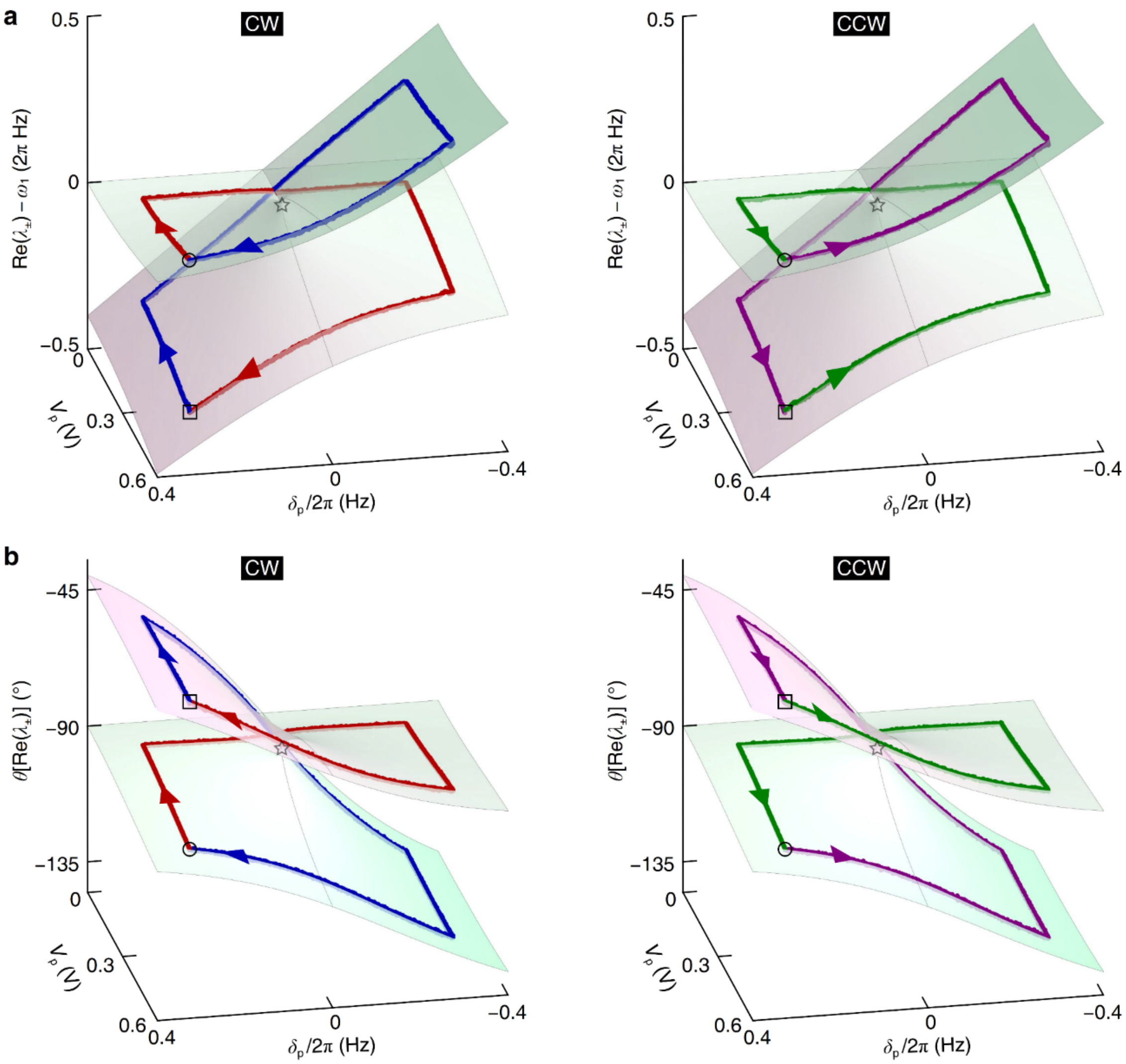

**Fig. 4. Results of smooth encircling starting from PT-symmetric phase obtained using the adaptive phase-tracking technique. a**, The phase-tracked closed-loop oscillation frequencies for the CW encircling process starting from the high-frequency sheet (red curve), the CW encircling process from the low-frequency sheet (blue curve), the CCW encircling process from the high-frequency sheet (purple curve), and the CCW encircling process from the low-frequency sheet (green curve) smoothly evolve on the $\mathrm{Re}[\lambda_{\pm}(t)]$ Riemann surface. The black circles and squares denote the start/end points of the encircling trajectories on the high and low-frequency sheets, respectively. The black star indicates the EP. **b**, The corresponding tracked phases for the four encircling processes. The colors on all the surfaces represent phase values.

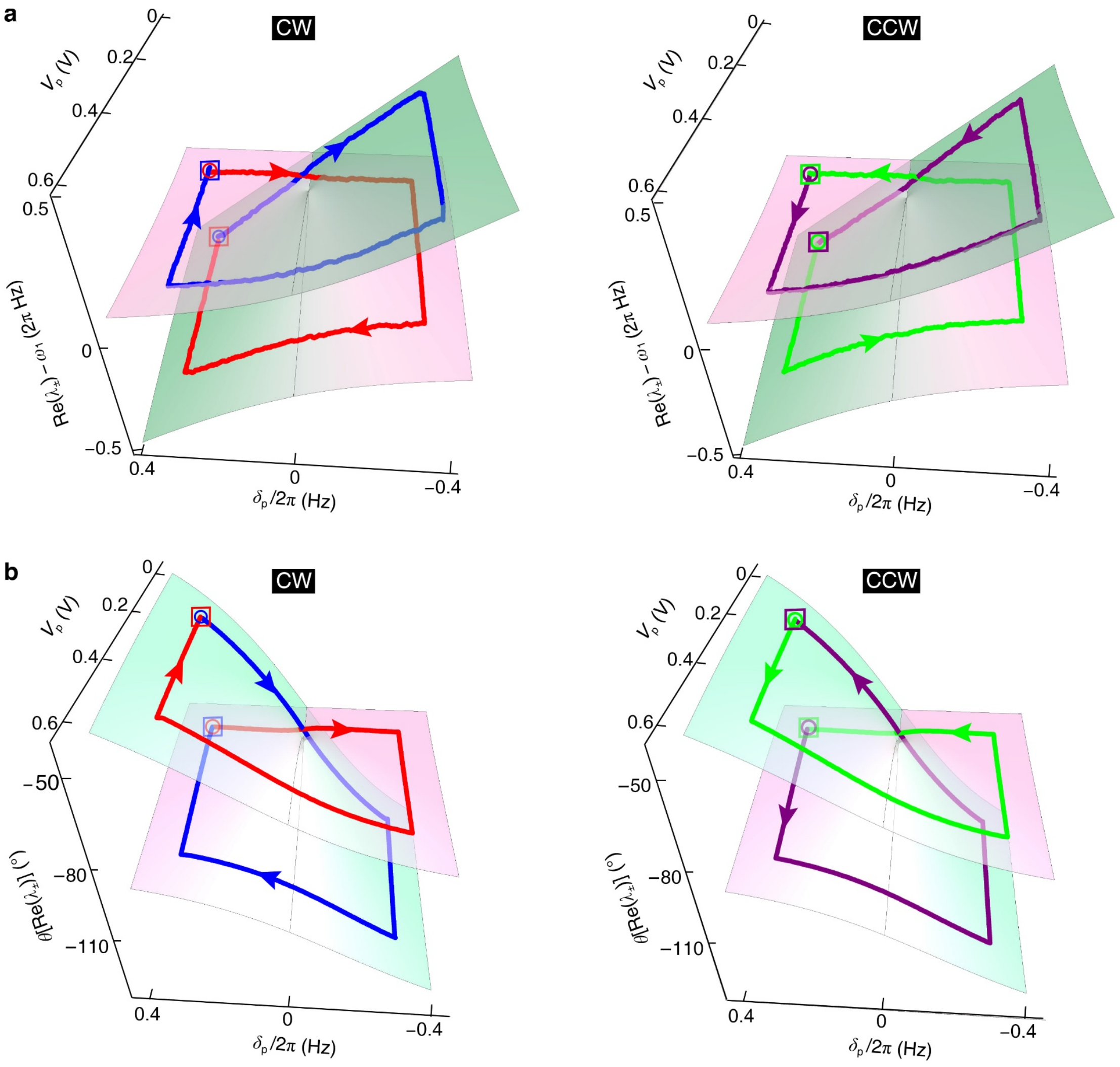

**Fig. 5. Results of smooth encircling starting from PT-broken phase obtained using the adaptive phase-tracking technique. a**, The phase-tracked closed-loop oscillation frequencies for the CW encircling process starting from the high-frequency sheet (red curve), the CW encircling process from the low-frequency sheet (blue curve), the CCW encircling process from the high-frequency sheet (purple curve), and the CCW encircling process from the low-frequency sheet (green curve) smoothly evolve on the Re$[\lambda_\pm(t)]$ Riemann surface. The circles and squares denote the start/end points of the encircling trajectories, respectively, while the red (purple) or blue (green) colors represent the CW (CCW) encircling starting on the high or low-frequency sheets, respectively. The arrow indicates the direction. **b**, The corresponding tracked phases for the four encircling processes. The colors on all the surfaces correspond to the imaginary part Im$(\lambda_\pm)$ of the eigenvalues.

# Dynamically encircling an exceptional point through phase-tracked closed-loop control

Sen Zhang[1], Yangyu Huang[1], Lei Yu[2], Kaixuan He[2], Ning Zhou[2],
Dingbang Xiao[1], Xuezhong Wu[1], Franco Nori[3,4], Hui Jing[5*] & Xin Zhou[1*]

[1]College of Intelligence Science and Technology, NUDT, Changsha, 410073, China.
[2]East China Institute of Photo-Electronic IC, Bengbu, 233042, China.
[3]Center for Quantum Computing (RQC), RIKEN, Wakoshi, Saitama, 351-0198, Japan.
[4]Department of Physics, University of Michigan, Ann Arbor, MI 48109-1040, USA.
[5]Key Laboratory of Low-Dimensional Quantum Structures and Quantum Control of Ministry of Education, Hunan Normal University, Changsha, 410081, China.
*e-mail: jinghui73@foxmail.com; zhouxin11@nudt.edu.cn.

**CONTENTS**

## SUPPLEMENTARY NOTE 1. DEVICE FABRICATION

As illustrated in Supplementary Fig. 7a, the fabrication of the device commences with a 6-inch silicon-on-insulator (SOI) wafer. This wafer comprises a 10 μm-thick P-type ⟨100⟩ single-crystal silicon conductive layer with a resistivity of 0.01 Ω·cm, a 3 μm-thick oxide layer, and a 500 μm-thick single-crystal silicon handle layer. The fabrication process involves several key steps:

**1. Platform etching:** Shallow cavities are etched into the conductive layer of the substrate SOI using a deep reactive ion etching (DRIE) technique to create bonding platforms and conductive routings. Importantly, the tops of the bonding platforms remain unetched, while the tops of the conductive routings are etched, achieving a height of 5 μm. Prior to this etching process, alignment marks are etched on the handle layer to facilitate subsequent patterning and bonding operations.

**2. Wafer direct bonding:** A structure SOI is then bonded to the etched substrate SOI through a direct wafer bonding process. This structure SOI consists of a 100 μm-thick P-type ⟨100⟩ single-crystal silicon structural layer, also exhibiting a resistivity of 0.01 Ω·cm, in addition to a 3 μm-thick oxide layer and a 450 μm-thick single-crystal silicon handle layer. The structure layer of the second SOI is tightly bonded to the platforms of the substrate SOI.

**3. Chemical mechanical polishing (CMP):** The handle and oxide layers of the structure SOI are subsequently removed using a CMP process, leaving a 100 μm-thick structural layer firmly bonded to the substrate SOI.

**4. Metal pads patterning:** With the aid of the previously etched alignment marks, metal pads are patterned and spurted onto the structural layer to facilitate wire bonding.

**5. Structure release:** The structure layer is patterned using a photolithography process and then etched through using a DRIE process to release the resonator, the electrodes, and the pads. The alignment of the structure with the platforms underneath is guaranteed using the alignment marks.

The completed wafer, as depicted in Supplementary Fig. 7b, contains a total of 333 devices. The dicing process employs stealth laser cutting along the dicing lanes shown in Supplementary Fig. 7c, during which the laser is focused inside the wafer to induce a series of defects. These internal imperfections serve as starting points for crack propagation. By applying tension to the tape adhered to the wafer's underside, the devices are separated from the cut lanes due to stress concentration. This technique avoids contamination of the bare devices, as the laser cutting occurs internally. Finally, the microscopic image of a fabricated non-Hermitian MEMS resonator is presented in Supplementary Fig. 7d, and the device is subsequently packaged in a carrier to maintain a vacuum of 0.001 Pa, thereby substantially mitigates air damping effects.

## SUPPLEMENTARY NOTE 2. EQUATIONS OF MOTION AND THE EFFECTIVE HAMILTONIAN

### A. Equations of motion in driving frames

We commence with the classical Newtonian equations of motion for the system, described by

$$\begin{bmatrix} \ddot{x} \\ \ddot{y} \end{bmatrix} + \begin{bmatrix} \gamma_1 & 0 \\ 0 & \gamma_2 \end{bmatrix} \begin{bmatrix} \dot{x} \\ \dot{y} \end{bmatrix} + \begin{bmatrix} \omega_1^2 + \Delta_\mathrm{p}/2 & \Delta_\mathrm{p}/2 \\ \Delta_\mathrm{p}/2 & \omega_2^2 + \Delta_\mathrm{p}/2 \end{bmatrix} \begin{bmatrix} x \\ y \end{bmatrix} = \begin{bmatrix} F\cos(\omega_\mathrm{d} t)/m \\ 0 \end{bmatrix}, \tag{S.1}$$

where $\Delta_\mathrm{p} = 2\kappa V_0 V_\mathrm{p} \cos(\omega_\mathrm{p} t) - 2\kappa V_\mathrm{p}^2/2$. We apply the rotating-wave approximation that considers higher-harmonic idler waves, the displacements can be expanded as

$$x = \sum_{m=-\infty}^{\infty} \frac{1}{2} A_m e^{-i(\omega_\mathrm{d} + m\omega_\mathrm{p})t} + \text{c.c.},$$
$$y = \sum_{m=-\infty}^{\infty} \frac{1}{2} B_m e^{-i(\omega_\mathrm{d} + m\omega_\mathrm{p})t} + \text{c.c.}, \tag{S.2}$$

where c.c. denotes complex conjugation, $A_m$ and $B_m$ are the slowly varying complex amplitudes of the $m$-th idler wave. The 0-th idler wave is the exact mode itself. Substituting (S.2) into (S.1) and omitting the second-order time derivatives of the complex amplitudes, we have,

$$\left[\gamma_1 - 2i(\omega_\mathrm{d} + m\omega_\mathrm{p})\right] \dot{A}_m - \left[(\omega_\mathrm{d} + m\omega_\mathrm{p})^2 - \omega_1^2 + \frac{\kappa V_\mathrm{p}^2}{4} + i\gamma_1(\omega_\mathrm{d} + m\omega_\mathrm{p})\right] A_m$$
$$-\frac{\kappa V_\mathrm{p}^2}{4} B_m + \frac{\kappa V_0 V_\mathrm{p}}{2}(A_{m-1} + A_{m+1}) + \frac{\kappa V_0 V_\mathrm{p}}{2}(B_{m-1} + B_{m+1}) = \frac{F}{2m}\delta_{m,0}, \tag{S.3}$$
$$\left[\gamma_2 - 2i(\omega_\mathrm{d} + m\omega_\mathrm{p})\right] \dot{B}_m - \left[(\omega_\mathrm{d} + m\omega_\mathrm{p})^2 - \omega_2^2 + \frac{\kappa V_\mathrm{p}^2}{4} + i\gamma_2(\omega_\mathrm{d} + m\omega_\mathrm{p})\right] B_m$$
$$-\frac{\kappa V_\mathrm{p}^2}{4} A_m + \frac{\kappa V_0 V_\mathrm{p}}{2}(B_{m-1} + B_{m+1}) + \frac{\kappa V_0 V_\mathrm{p}}{2}(A_{m-1} + A_{m+1}) = 0. \tag{S.4}$$

In this study, actuation is applied to mode 1, $\omega_\mathrm{d} \sim \omega_1$. The pump frequency is tuned close to the frequency difference, $\omega_\mathrm{p} \sim \Delta\omega \equiv \omega_2 - \omega_1$. Consequently, this configuration facilitates the up-conversion of the drive and pump, which effectively reaches mode 2, $\omega_\mathrm{d} + \omega_\mathrm{p} \sim \omega_2$. To simplify our analysis, we focus exclusively on the first-order dynamical interaction, which is captured by setting $m = 0$ in (S.3) and $m = 1$ in (S.4),

$$(\gamma_1 - 2i\omega_\mathrm{d}) \dot{A}_m - \left(\omega_\mathrm{d}^2 - \omega_1^2 + \frac{\kappa V_\mathrm{p}^2}{4} + i\gamma_1\omega_\mathrm{d}\right) A_0 + \frac{\kappa V_0 V_\mathrm{p}}{2} B_1 = \frac{F}{2m}, \tag{S.5}$$

$$\left[\gamma_2 - 2i(\omega_\mathrm{d} + \omega_\mathrm{p})\right] \dot{B}_1 - \left[(\omega_\mathrm{d} + \omega_\mathrm{p})^2 - \omega_2^2 + \frac{\kappa V_\mathrm{p}^2}{4} + i\gamma_2(\omega_\mathrm{d} + \omega_\mathrm{p})\right] B_1$$
$$+ \frac{\kappa V_0 V_\mathrm{p}}{2} A_0 = 0. \tag{S.6}$$

By dividing equations (S.5) and (S.6) by $2\omega_1$ and $2\omega_2$, respectively, we derive the approximated first-order equations of motion within the driving rotating frames,

$$i\begin{bmatrix}\dot{A}_0\\ \dot{B}_1\end{bmatrix}=\begin{bmatrix}\omega_1-\frac{\kappa V_{\rm p}^2}{8\omega_1}-\omega_{\rm d}-i\gamma_1/2 & \frac{\kappa V_0 V_{\rm p}}{4\omega_1}\\ \frac{\kappa V_0 V_{\rm p}}{4\omega_2} & \omega_2-\frac{\kappa V_{\rm p}^2}{8\omega_2}-\omega_{\rm p}-\omega_{\rm d}-i\gamma_2/2\end{bmatrix}\begin{bmatrix}A_0\\ B_1\end{bmatrix}-\begin{bmatrix}\frac{F}{4m\omega_1}\\ 0\end{bmatrix}.$$

During the calculations, we employ several approximations: $\gamma_{1,2}/(2\omega_{1,2})\sim 0$, $\omega_{\rm d}\sim\omega_1$, $\omega_{\rm d}+\omega_{\rm p}\sim\omega_2$, $(\omega_{\rm d}^2-\omega_1^2)/(2\omega_1)\sim\omega_{\rm d}-\omega_1$, and $[(\omega_{\rm d}+\omega_{\rm p})^2-\omega_2^2]/(2\omega_2)\sim\omega_{\rm d}+\omega_{\rm p}-\omega_2$. Updating $B_1$ by a slightly scaled variable $B_1\omega_2/\omega_1$, we can further obtain the formal first-order equations of motion for the system in the driving frames,

$$i\begin{bmatrix}\dot{A}_0\\ \dot{B}_1\end{bmatrix}=\begin{bmatrix}\Omega_1-\omega_{\rm d}-i\gamma_1/2 & g\\ g & \Omega_2-\omega_{\rm d}-i\gamma_2/2\end{bmatrix}\begin{bmatrix}A_0\\ B_1\end{bmatrix}-\begin{bmatrix}f\\ 0\end{bmatrix}, \tag{S.7}$$

where, $\Omega_1=\omega_1-\kappa V_{\rm p}^2/(8\omega_1)$, $\Omega_2=\omega_2-\kappa V_{\rm p}^2/(8\omega_2)-\omega_{\rm p}$, $g=\kappa V_0 V_{\rm p}/(4\omega_1)$, and $f=F/(4m\omega_1)$.

### B. Effective Hamiltonian

The equations of motion (S.7) are formulated in fast rotating driving frames of $\omega_{\rm d}$ for mode 1 and $\omega_{\rm d}+\omega_{\rm p}$ for mode 2. To facilitate analysis, we now transform these equations into a slowly rotating frame of $\omega_{\rm p}$, which is known as the Floquet frame. This transformation is achieved by letting

$$x_0=\frac{1}{2}A_0e^{-i\omega_{\rm d}t}+\text{c.c.},$$
$$y_1=\frac{1}{2}B_1e^{-i\omega_{\rm d}t}+\text{c.c.},$$

or

$$A_0=2x_0e^{i\omega_{\rm d}t}-A_0^*e^{2i\omega_{\rm d}t},$$
$$B_1=2y_1e^{i\omega_{\rm d}t}-B_1^*e^{2i\omega_{\rm d}t}. \tag{S.8}$$

where the superscript "*" denotes the complex conjugation of the variable. Substituting (S.8) into (S.7), we have

$$\begin{aligned}&2i(\dot{x}_0+i\omega_{\rm d}x_0)e^{i\omega_{\rm d}t}-i(\dot{A}_0^*+2i\omega_{\rm d}A_0^*)e^{2i\omega_{\rm d}t}\\ &=\left(\Omega_1-\omega_{\rm d}-i\frac{\gamma_1}{2}\right)\left(2x_0e^{i\omega_{\rm d}t}-A_0^*e^{2i\omega_{\rm d}t}\right)+g\left(2y_1e^{i\omega_{\rm d}t}-B_1^*e^{2i\omega_{\rm d}t}\right)\\ &-2f\cos(\omega_{\rm d}t)e^{i\omega_{\rm d}t}+fe^{2i\omega_{\rm d}t},\end{aligned} \tag{S.9}$$

$$\begin{aligned}&2i(\dot{y}_1+i\omega_{\rm d}y_1)e^{i\omega_{\rm d}t}-i(\dot{B}_1^*+2i\omega_{\rm d}B_1^*)e^{2i\omega_{\rm d}t}\\ &=\left(\Omega_2-\omega_{\rm d}-i\frac{\gamma_2}{2}\right)\left(2y_1e^{i\omega_{\rm d}t}-B_1^*e^{2i\omega_{\rm d}t}\right)+g\left(2x_0e^{i\omega_{\rm d}t}-A_0^*e^{2i\omega_{\rm d}t}\right)\end{aligned} \tag{S.10}$$

Equating the coefficients of like power of $e^{i\omega_\mathrm{d}t}$ in (S.9) and (S.10), we have

$$i\dot{x}_0 = \left(\Omega_1 - i\frac{\gamma_1}{2}\right)x_0 + gy_1 - f\cos(\omega_\mathrm{d}t), \tag{S.11}$$

$$i\dot{A}_0^* = \left(\Omega_1 + \omega_\mathrm{d} - i\frac{\gamma_1}{2}\right)A_0^* + gB_1^* - f, \tag{S.12}$$

and

$$i\dot{y}_1 = \left(\Omega_2 - i\frac{\gamma_2}{2}\right)y_1 + gx_0, \tag{S.13}$$

$$i\dot{B}_1^* = \left(\Omega_2 + \omega_\mathrm{d} - i\frac{\gamma_2}{2}\right)B_1^* + gA_0^*, \tag{S.14}$$

Equations (S.12) and (S.14) are the equations of motion in the negative-frequency frames, which are conjugations of (S.7). Equations (S.11) and (S.13) constitute the equations of motion in the Floquet frame,

$$i\begin{bmatrix}\dot{x}_0\\ \dot{y}_1\end{bmatrix} = \begin{bmatrix}\Omega_1 - i\gamma_1/2 & g\\ g & \Omega_2 - i\gamma_2/2\end{bmatrix}\begin{bmatrix}x_0\\ y_1\end{bmatrix} - \begin{bmatrix}f\cos(\omega_\mathrm{d}t)\\ 0\end{bmatrix}. \tag{S.15}$$

The dynamic matrix of the equations of motion (S.15) is regarded as the effective Hamiltonian,

$$\mathbf{H} = \begin{bmatrix}\Omega_1 - i\frac{\gamma_1}{2} & g\\ g & \Omega_2 - i\frac{\gamma_2}{2}\end{bmatrix} = \begin{bmatrix}\omega_1 - \frac{\kappa V_\mathrm{p}^2}{8\omega_1} - i\frac{\gamma_1}{2} & g\\ g & \omega_1 - \frac{\kappa V_\mathrm{p}^2}{8\omega_2} - \delta_\mathrm{p} - i\frac{\gamma_2}{2}\end{bmatrix}. \tag{S.16}$$

where $\delta_\mathrm{p}$ is the detuning of the pump, $\delta_\mathrm{p} \equiv \omega_\mathrm{p} - \Delta\omega$. The eigenvalues of this non-Hermitian Hamiltonian is obtained by calculating $\det(\mathbf{H} - \lambda\mathbf{I}) = 0$,

$$\begin{aligned}\lambda_\pm =& \frac{\Omega_1 + \Omega_2}{2} - i\frac{\gamma_1 + \gamma_2}{4} \pm \sqrt{\left(\frac{\Omega_2 - \Omega_1}{2} - i\frac{\gamma_2 - \gamma_1}{4}\right)^2 + g^2}\\ =& \omega_1 - \frac{\kappa V_\mathrm{p}^2}{16}\left(\frac{1}{\omega_1} + \frac{1}{\omega_2}\right) - i\frac{\gamma_1 + \gamma_2}{4}\\ & \pm \sqrt{\left[\frac{\kappa V_\mathrm{p}^2}{16}\left(\frac{1}{\omega_1} - \frac{1}{\omega_2}\right) - \frac{\delta_\mathrm{p}}{2} - i\frac{\gamma_2 - \gamma_1}{4}\right]^2 + g^2}.\end{aligned} \tag{S.17}$$

### C. Steady-state responses

The steady-state information of the system can be obtained by letting $\dot{A}_0 = \dot{B}_1 = 0$ for (S.7), which gives

$$\begin{bmatrix}A_0\\ B_1\end{bmatrix} = \begin{bmatrix}\Omega_1 - \omega_\mathrm{d} - i\gamma_1/2 & g\\ g & \Omega_2 - \omega_\mathrm{d} - i\gamma_2/2\end{bmatrix}^{-1}\begin{bmatrix}f\\ 0\end{bmatrix}, \tag{S.18}$$

The complex amplitudes of the mode 1 and the first idler wave of mode 2 are given by

$$A_0 = f\frac{\Omega_2 - \omega_{\mathrm{d}} - i\gamma_2/2}{(\omega_{\mathrm{d}} - \lambda_+)(\omega_{\mathrm{d}} - \lambda_-)}, \tag{S.19}$$

$$B_1 = f\frac{-g}{(\omega_{\mathrm{d}} - \lambda_+)(\omega_{\mathrm{d}} - \lambda_-)}. \tag{S.20}$$

The theoretical amplitude responses are obtained by calculating $|A_0|$ and $|B_1|$. The phase response of mode 1 is given by $-\mathrm{Arg}(A_0)$. It is noteworthy that the phase of the mode-2 idler wave is not directly represented by $-\mathrm{Arg}(B_1)$ due to the contribution of the pump phase to the overall phase of the idler wave. Since the pump phase is arbitrary and uncontrollable in this study, the overall phase of the mode-2 idler wave is undetectable.

**SUPPLEMENTARY FIGURES.**

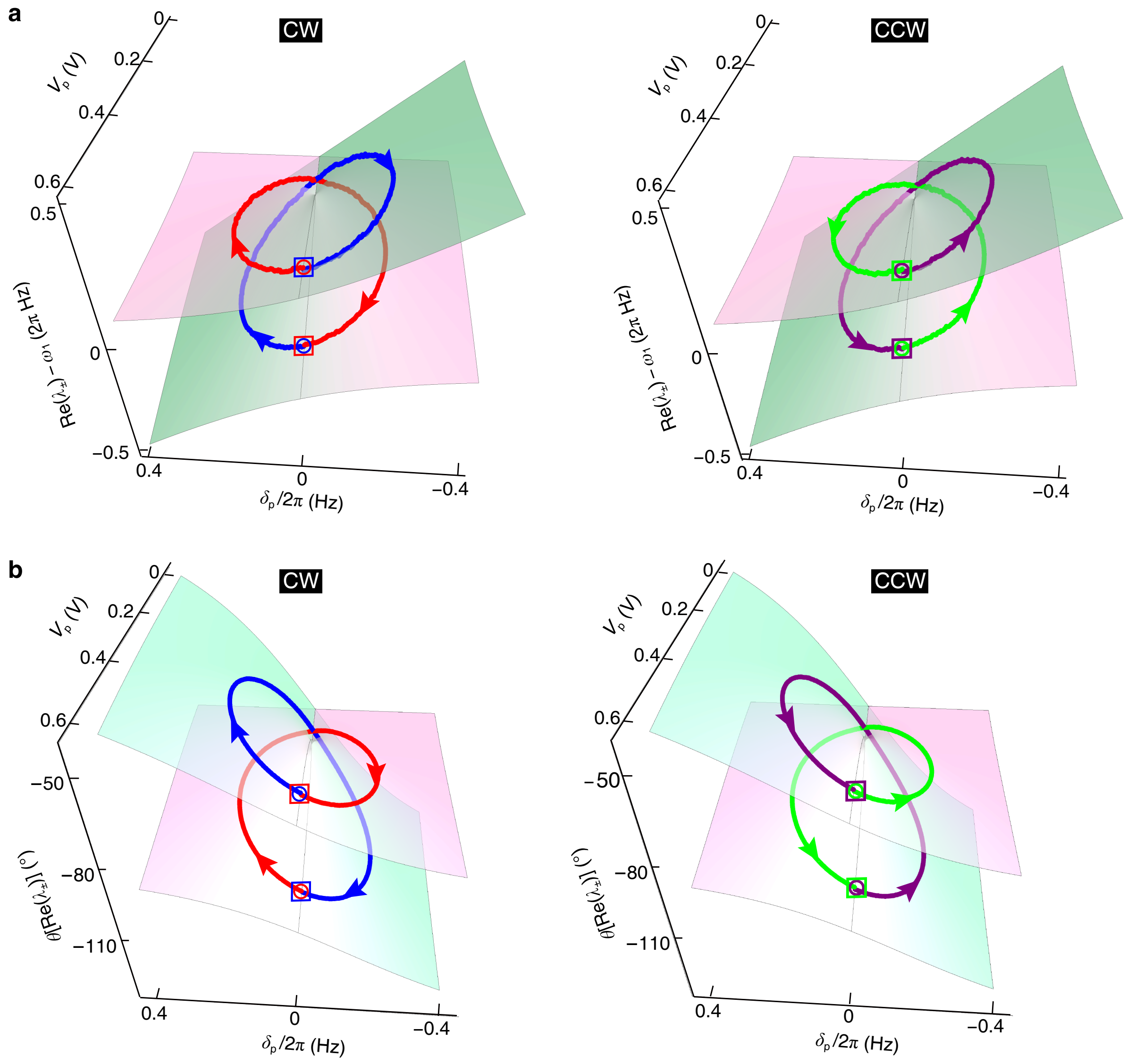

**Supplementary Figure 1**. **Results of the smoothly encircling of EP with a circular trajectory.** (**a**) The phase-tracked closed-loop oscillation frequencies for the CW encircling process starting from the high-frequency sheet (red curve), the CW encircling process from the low-frequency sheet (blue curve), the CCW encircling process from the high-frequency sheet (purple curve), and the CCW encircling process from the low-frequency sheet (green curve) smoothly evolve on the Re($\lambda_\pm$) Riemann surface. The circles and squares denote the start/end points of the encircling trajectories on the high and low-frequency sheets, respectively. The arrow indicates the direction. (**b**) The corresponding tracked phases for the four encircling processes. The colors on all the surfaces correspond to the imaginary part Im($\lambda_\pm$) of the eigenvalues.

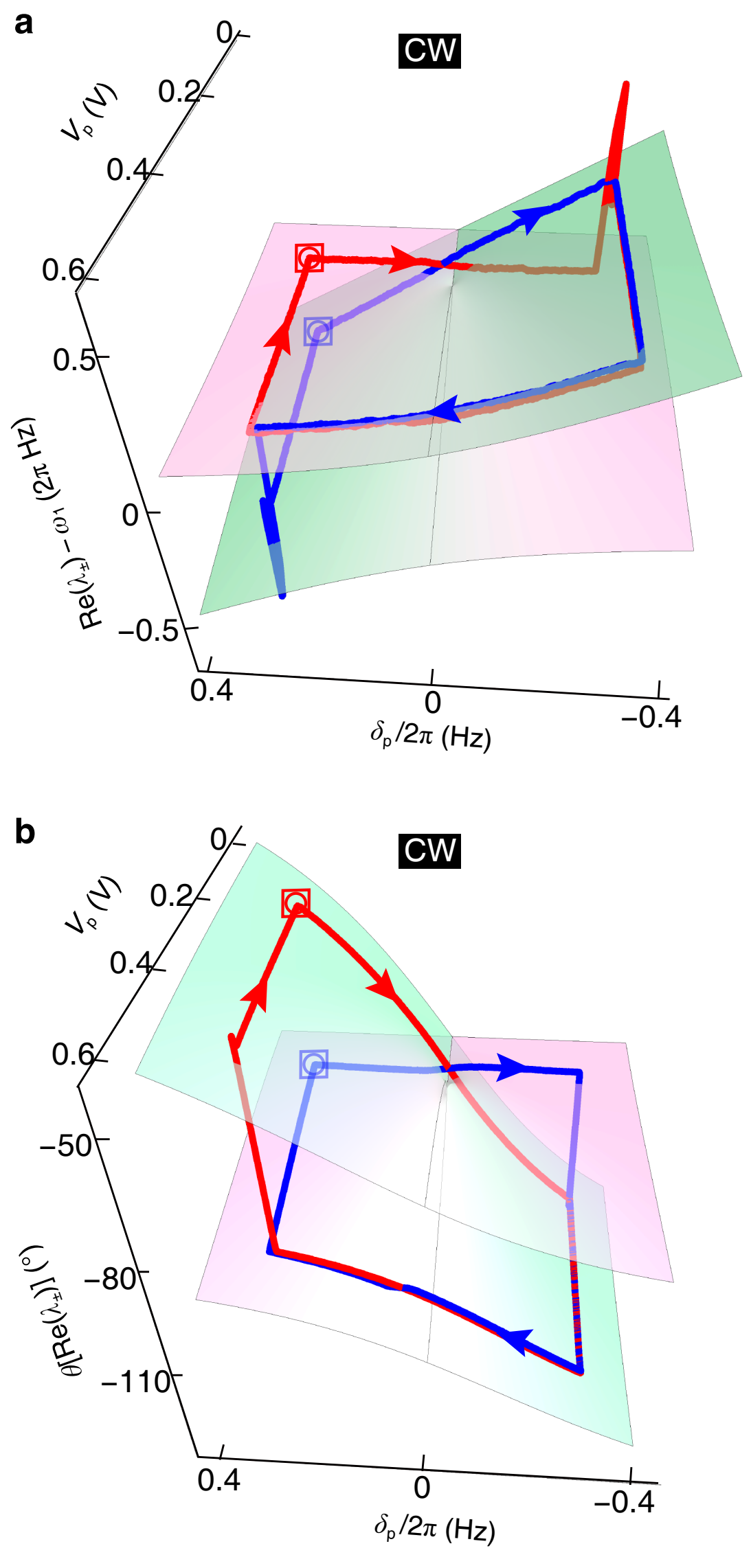

**Supplementary Figure 2**. **Results of non-adiabatic transitions through phase-tracked closed-loop control.** (**a**) The phase-tracked closed-loop oscillation frequencies for the CW encircling process starting from the high-frequency sheet (red curve), the CW encircling process from the low-frequency sheet (blue curve), the CCW encircling process from the high-frequency sheet (purple curve), and the CCW encircling process from the low-frequency sheet (green curve) exhibit a transition between the high- and low-frequency sheets of the Riemann surface in lower-loss states. The circles and squares denote the start/end points of the encircling trajectories on the high and low-frequency sheets, respectively. The arrow indicates the direction. (**b**) The corresponding tracked phases for the four encircling processes. The colors on all the surfaces represent the system's energy dissipation, with pink indicating low-loss states and cyan representing high-loss states.

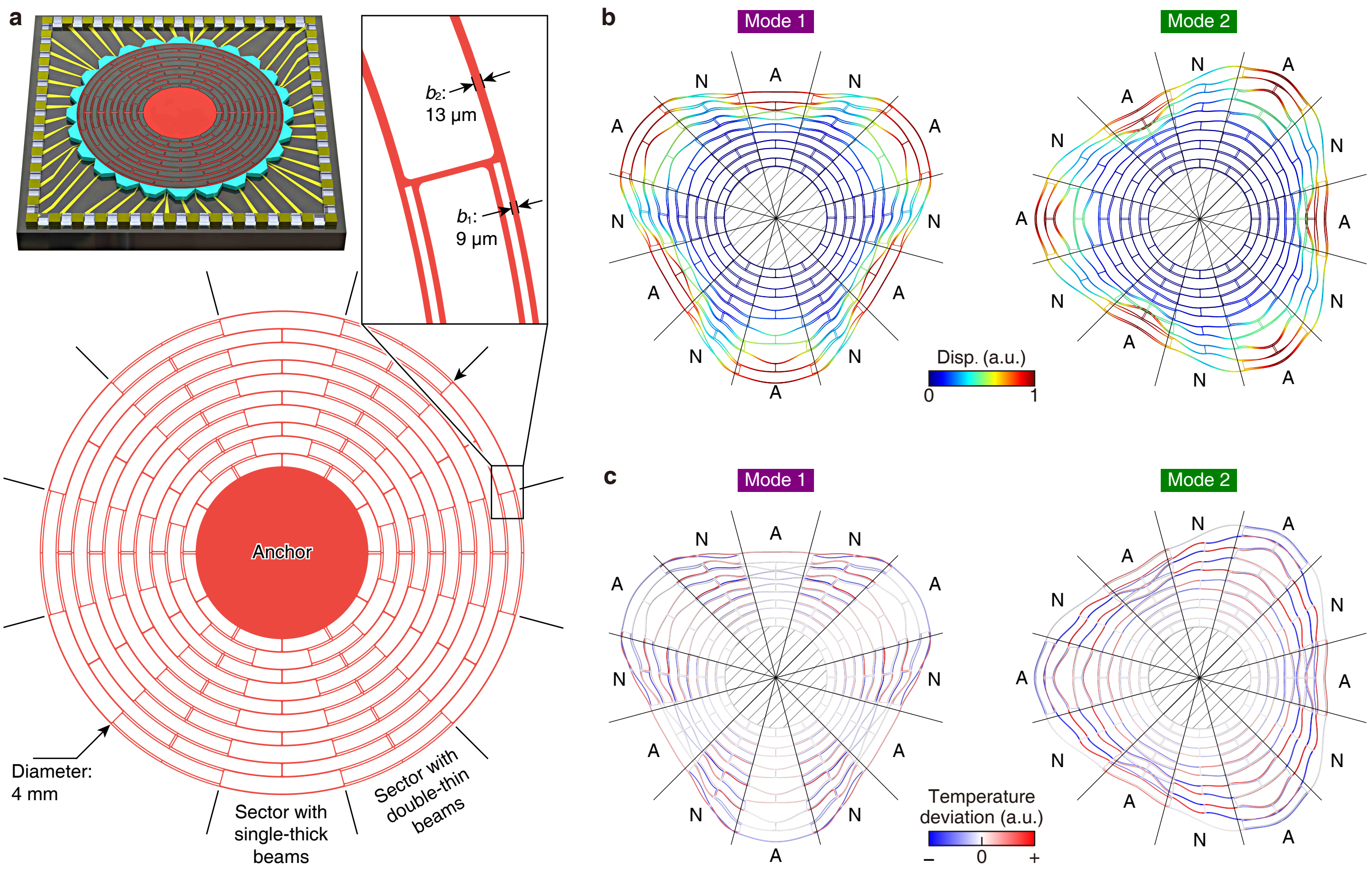

**Supplementary Figure 3**. **Device Design.** (**a**) Structure of the non-Hermitian disk resonator, highlighted in red. The capacitive electrodes are indicated in cyan, while the electric routings and pads are marked in yellow. The substrate is shown in gray. The resonator has a diameter of 4 mm and a height of 100 $\mu$m, with a capacitive gap of 9 $\mu$m. The widths of the thin and thick beams are 9 $\mu$m and 13 $\mu$m, respectively. (**b**) The six-node in-plane standing-wave modes displaying normalized displacements. (**c**) The instantaneous temperature-deviation fields resulting from thermal-elastic coupling.

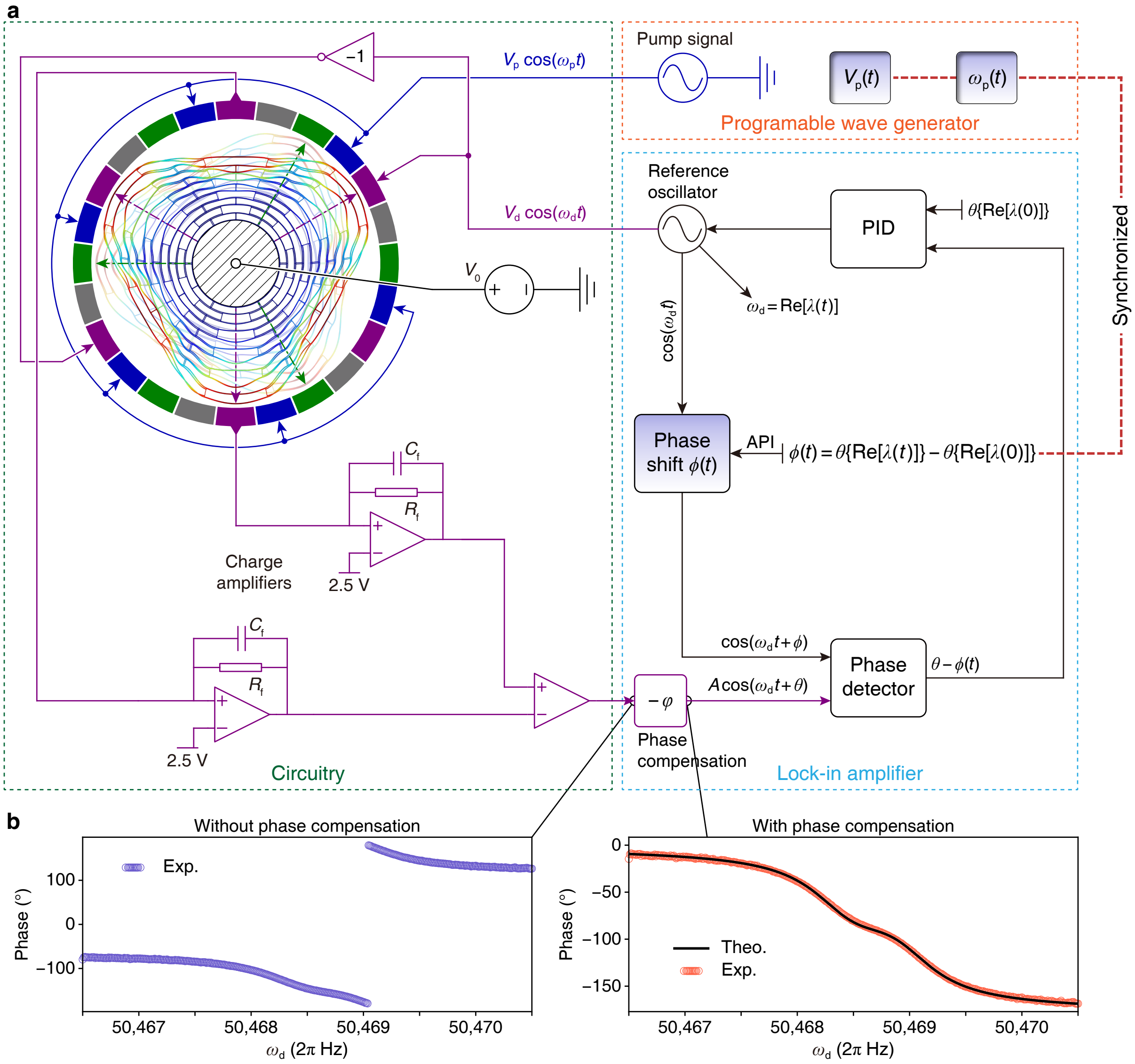

**Supplementary Figure 4**. **Detailed experimental setup.** (**a**) Mode 1 is actuated differentially, and its response displacement is measured capacitively. A parametric pump with adjustable amplitude and frequency is employed to modulate the coupling stiffness. An adaptive PLL is implemented within a Zurich Instruments MFLI lock-in amplifier to track the phase $\theta\{\text{Re}[\lambda_\pm(t)]\}$ in real-time. The wave generator and the lock-in amplifier are synchronized to facilitate the coordinated variations of $V_\text{p}(t)$, $\delta_\text{p}(t)$, and $\phi(t)$. (**b**) The test circuitry can introduce an additional phase $\varphi = -65°$ to the output signal. This phase shift is compensated before the output signal is put into the phase detector. The left and right panels illustrate the phases of the open-loop output signal without and with phase compensation, respectively. The black curve represents the theoretically expected response phase, given by $\theta = -\text{Arg}(\chi_1)$.

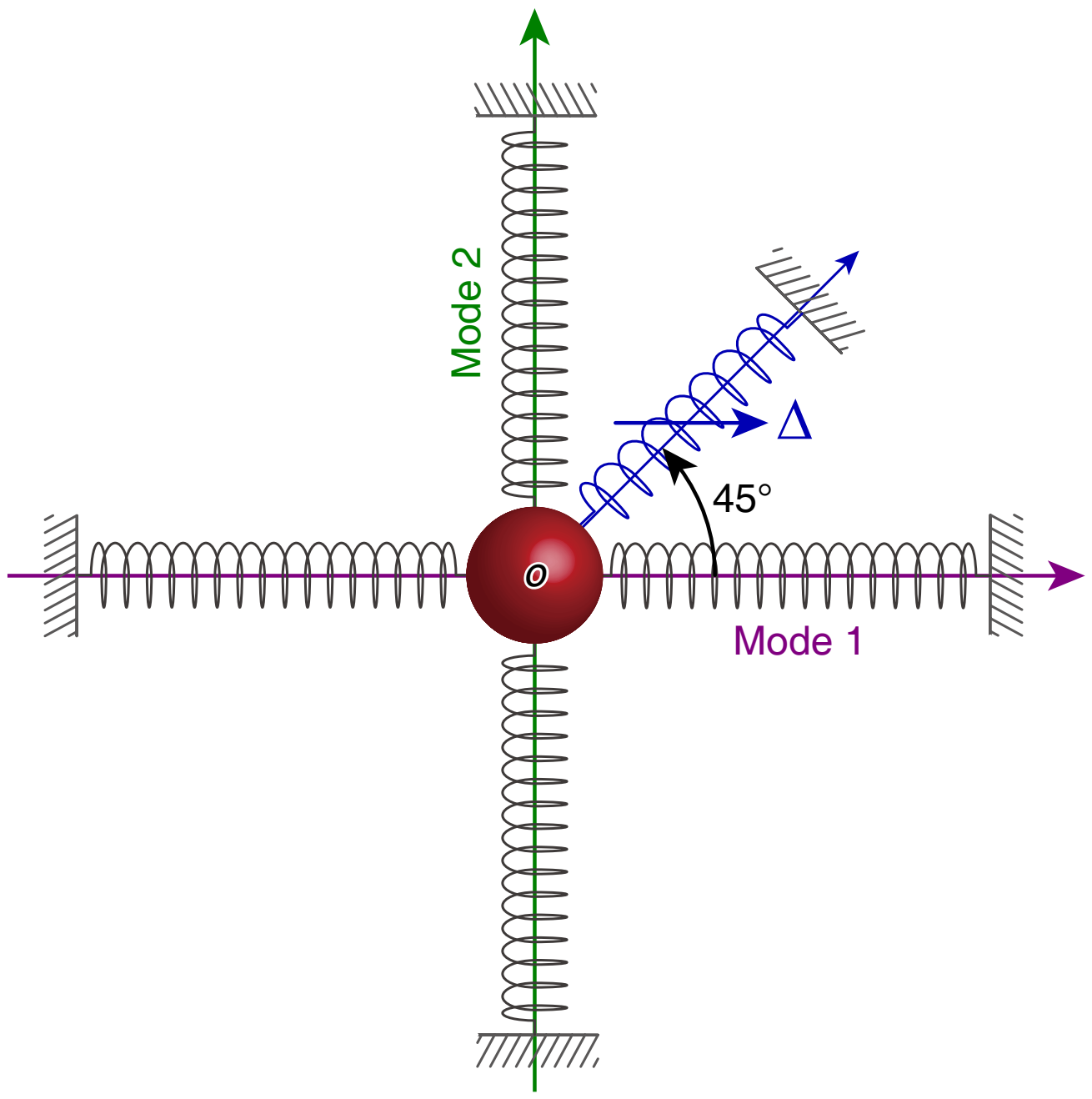

**Supplementary Figure 5**. **Reduced-order model of the dynamical coupling of the two operational modes.** Modes 1 and 2 can be represented as two orthogonal degrees of freedom of a single proof mass. The parametric pump is equivalent to the Floquet dynamical modulation of a spring oriented at a 45° off-axis angle.

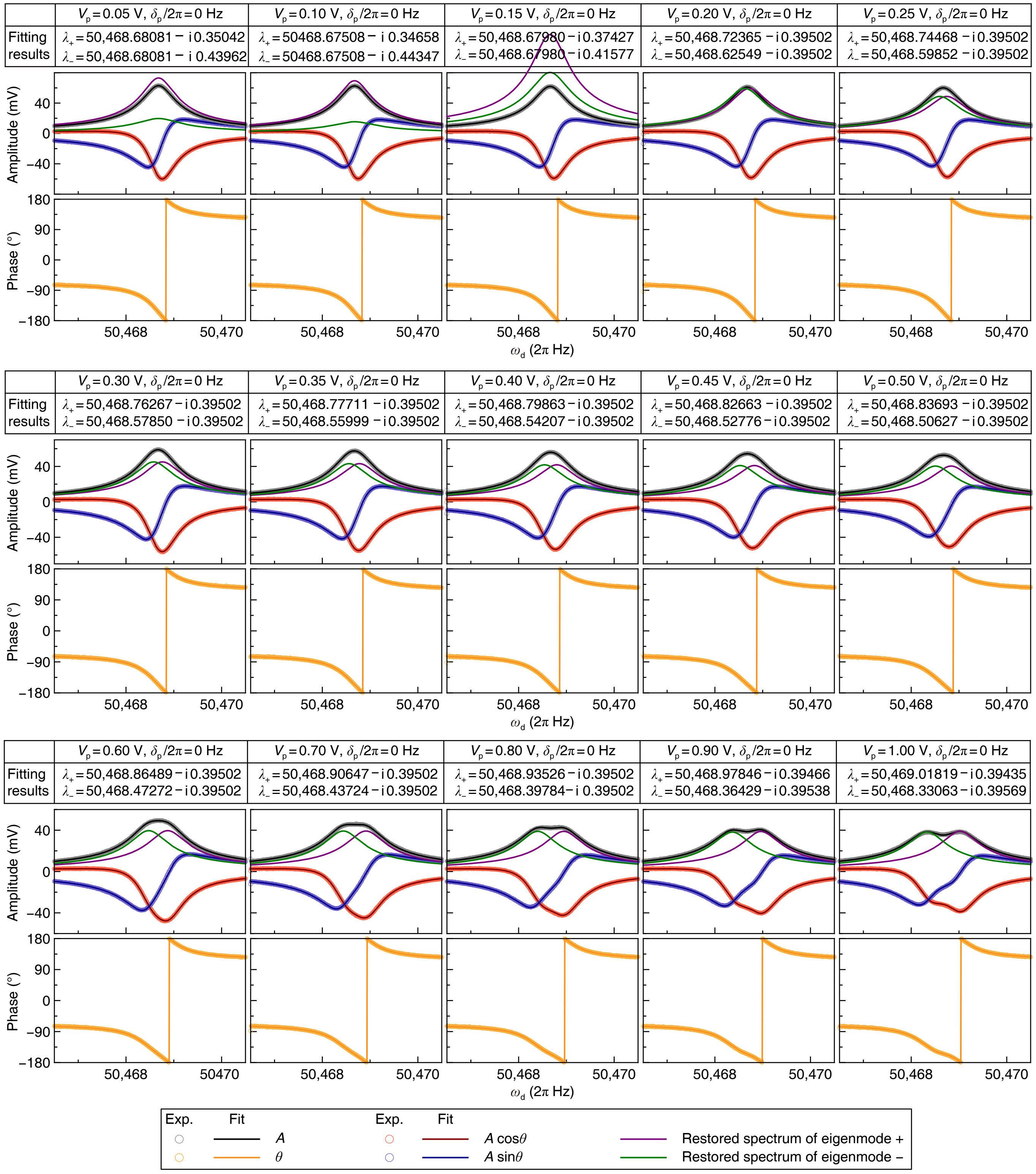

**Supplementary Figure 6**. **Fitting processes of the frequency responses to obtain the complex eigenvalues.** The in-phase and quadrature components of the measured data are fitted to the real and imaginary parts of the modified susceptibility, respectively, to extract the complex eigenvalues. Exp., experimental results. Fit., fitting results.

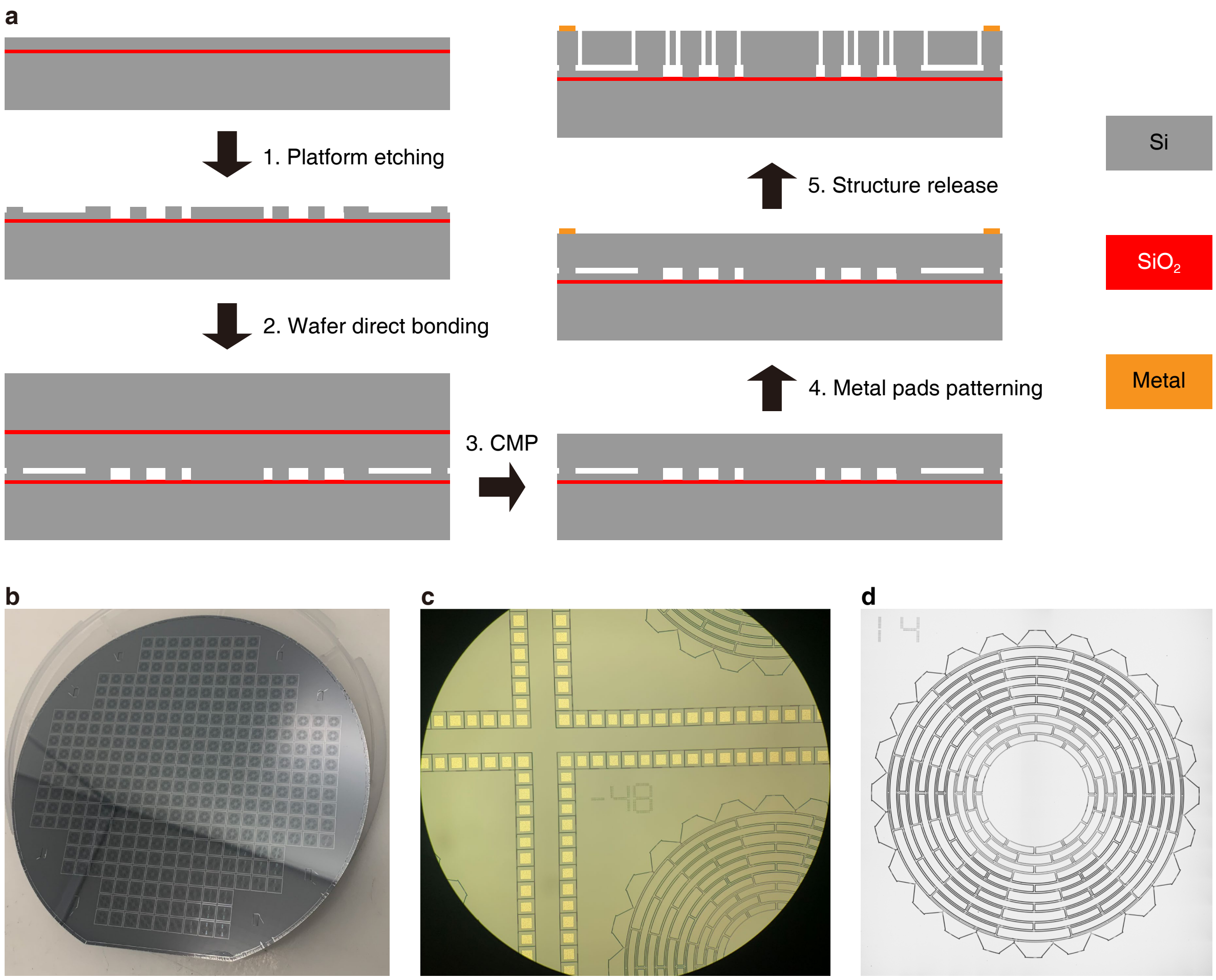

**Supplementary Figure 7**. **Fabrication of the non-Hermitian MEMS resonator.** (**a**) MEMS fabrication process of the devices. (**b**) The fabricated 6-inch wafer with 333 devices. (**c**) Details of the dicing lanes. (**d**) The microscopic picture of a non-Hermitian MEMS disk resonator.